\documentclass[12pt,epsfig,amsmath,amssymb,pstricks]{article}
\usepackage{graphicx}

\begin{document}

\title{\bf Decompositional equivalence:\\A fundamental symmetry underlying quantum theory}

\author{{Chris Fields}\\ \\
{\it 243 West Spain Street}\\
{\it Sonoma, CA 95476 USA}\\ \\
{fieldsres@gmail.com}}
\maketitle

\begin{abstract}
Decompositional equivalence is the principle that there is no preferred decomposition of the universe into subsystems.  It is shown here, by using a simple thought experiment, that quantum theory follows from decompositional equivalence together with Landauer's principle.  This demonstration raises within physics a question previously left to psychology: how do human - or any - observers identify or agree about what constitutes a ``system of interest''?
\end{abstract}


\newpage

\begin{quote}
``Nature is relentless and unchangeable, and it is indifferent as to whether its hidden reasons and actions are understandable to man or not.''
\begin{flushright}
-- Galileo Galilie
\end{flushright} 
\end{quote}

\section{Introduction}

The enormous empirical success of quantum theory provides strong evidence that quantum theory is true, at least in the pragmatic sense of enabling correct predictions about the observable behavior of the world.  What has been unclear for almost 90 years now is \textit{why} it is true.  Despite decades of effort, we have not found simple, intuitively-compelling physical principles that lead us to \textit{expect} quantum theory to be true.  The theory is, instead, almost universally regarded and presented to students as simply a mathematical framework that makes correct if counter-intuitive predictions.  Without a clear physical picture to motivate the mathematical formalism of quantum theory, we are left with a choice between weakly-constrained metaphysical speculations and ``shut up and calculate.''\footnote{The mathematical formalism of quantum theory has been subjected to physical and philosophical interpretation since its inception.  Bacciagaluppi and Valentini (2009) discuss the interpretative positions advanced by the theory's founders at the 1927 Solvay Conference and reproduce their original papers.  Bunge (1956) reviews the largely-unchanged interpretative landscape 30 years later.  Bastin (1971) provides a revealing glimpse of interpretative discussions following the introduction of Everett's (1957) relative-state interpretation, but prior to both the reformulation of Everett's interpretation in terms of ``multiple worlds'' by DeWitt (1970) and the introduction of decoherence by Zeh (1970).  Landsman (2007) and Wallace (2008) provide more recent synoptic reviews, the former with an emphasis on decoherence and the latter with an emphasis on multiple worlds.  The diversity of opinions on basic questions of interpretation remains large, as documented by Norsen and Nelson (2013), Schlosshauer, Kofler and Zeilinger (2013) and Sommer (2013) by surveying participants at relevant conferences.  Both physicists and philosophers have found the seemingly irresolvable differences between interpretative stances disturbing.  Fuchs (2002) parodies interpretative ``camps'' as fundamentalist churches.  Cabello (2015) titles a recent, fairly exhaustive overview of the diversity of interpretative assumptions a ``map of madness.''}

My goal in this paper is to show that standard, unitary quantum theory is materially implied by the conjunction of two simple, intuitively compelling physical principles.  The first of these is Landauer's (1961, 1999) principle that ``information is physical'' or more precisely: 1) classical information -- information encodable by strings of bits -- only ``exists'' when it has been encoded by a thermodynamically-irreversible state change in some physical system, and 2) any such encoding requires a finite expenditure of free energy (for further discussion, see Bennett, 2003).  The second principle is that all fundamental physical interactions are entirely invariant under arbitrary decompositions of any physical system into subsystems.  This principle of ``decompositional equivalence'' has been assumed in one form or another, generally implicitly, from the very beginnings of modern science; indeed the possibility of scientific investigation would seem to require a universe in which what we humans choose to call things -- in particular, what we choose to designate as the ``system of interest'' -- does not affect what is going on.  What is shown here is that these two principles, taken together, require formal descriptions of the states of physical systems that are inferred from the outcomes of finite observations to have the form specified by standard, unitary quantum theory.  Quantum theory thus emerges, in this treatment, not as a theory with a ``measurement problem,'' but as a clear and physically well-motivated formalism that precisely limits the information that any observer of any system can obtain by measurement.

The paper is organized as follows.  The next section defines decompositional equivalence, discusses its close relationship to classical reductionism, and shows that it is incorporated implicitly into the state-space formalisms employed by both classical and quantum physics.  Both classical and quantum physics, therefore, materially imply decompositional equivalence.  The following two sections discuss two consequences of decompositional equivalence: 1) there can be no distinct or ``preferred'' class of observers and 2) all finite observation -- all observation subject to Landauer's principle -- can be described as interactions between an observer and a \textit{black box} as defined by classical cybernetics, i.e. a system with observable external behavior but an unobservable and hence unknowable interior (Ashby, 1956; Moore, 1956).  The fifth section employs a simple thought experiment to show that if Landauer's principle is assumed, the state descriptions inferred from the outcomes of finite interactions between an observer and a black box must be those specified by standard, unitary quantum theory.  It shows, in particular, that the three principal axioms of quantum theory, those specifying the Hilbert space formalism, unitary evolution, and measurement by projection are materially implied by decompositional equivalence when a finite energetic cost of encoding observational outcomes is assumed.  The sixth section discusses the physical meaning of quantum superposition and shows how the standard quantum no-go theorems are entailed by decompositional equivalence plus Landauer's principle.  It shows, in particular, that superpositions of quantum \textit{states} can be replaced, without altering the formalism, by superpositions of \textit{systems}, even systems occupying classical states.  The seventh section illustrates the previous results using the double-slit experiment as an example.  The concluding section briefly discusses common assumptions of the ``classical worldview'' that violate decompositional equivalence, Landauer's principle or both and therefore contradict quantum theory.  These include assumptions that the sources of observational outcomes can be precisely identified and that system preparation or any other physical process involves local causation only.

\section{Decompositional equivalence: A ubiquitous but largely implicit assumption}

Physics -- indeed, all of science -- is motivated by a deep intuition that \textit{there is a way that the world works} and that this way that the world works is independent of what we humans or any other observers may say, do, or believe.  This intuition underlies both the idea that Nature has laws and the idea that those laws can be given a universal, observer-independent mathematical formulation.  It also underlies the idea of an experiment as a ``question to the world'' to which the world gives an unbiased answer.   

This deep intuition can be made more precise by noting that to conduct an experiment, an observer must: 1) designate a ``system of interest'' on which the experiment is to be performed, 2) choose the experiment to perform, and then 3) manipulate the system of interest in some way to ``prepare'' it for observation.  While the outcome of the experiment clearly depends on these three actions by the observer, it is universally assumed that the \textit{processes by which the world produces that outcome} do not.  These processes reflect how the world works, and how the world works is independent of what observers say, do or believe.  This distinction between the overt, observable behavior of the world and, as Galileo put it, the ``hidden reasons and actions'' that produce that behavior is a key component of the intuition that there is a way that the world works.  Consistent with common usage, these hidden reasons and actions can be referred to as ``fundamental interactions'' and taken, for the purposes of the present argument, to be the fundamental interactions postulated by the Standard Model plus gravity, string theory, M theory, or some future fundamental physics.

Let us focus on the first of the actions that an observer must take to perform an experiment: the designation of a system of interest.  A second deep intuition in both physics and other sciences is that there are no \textit{principled} limitations on what can be subjected to scientific investigation.  At least in principle, the choice of a system of interest is \textit{entirely arbitrary}; any system whatsoever can be chosen.   This arbitrariness is nicely illustrated by the common practice of considering some arbitrary voxel when discussing classical fluid flow; the voxel boundaries can be drawn anywhere, without restriction, to focus attention on the behavior of some small sample of the fluid.  The arbitrary movability of the ``Heisenberg cut'' in Copenhagen quantum theory provides another example.  Conceptually including the measurement apparatus or even the observer's sensory organs within the ``system of interest'' makes no difference to the way that the world works and hence no difference to the processes that produce the experimental outcome.

The principle of decompositional equivalence combines these two deep intuitions by noting that the choice of a system of interest is an instance of \textit{system decomposition}; in particular, it is the decomposition of ``the world'' into the system of interest and ``everything else.''  The principle generalizes this instance to all decompositions, by stating:
\begin{quote}
\textbf{Decompositional equivalence}:  All fundamental physical interactions are entirely invariant under arbitrary decompositions of any physical system into subsystems.
\end{quote}
In particular, all fundamental physical interactions are entirely invariant under arbitrary decompositions of the universe $\mathbf{U}$ into a system of interest $\mathbf{S}$, an observer $\mathbf{O}$ and a surrounding environment $\mathbf{E}$, where $\mathbf{E}$ is taken to include everything in $\mathbf{U}$ not included in either $\mathbf{S}$ or $\mathbf{O}$.  When stated in this way, decompositional equivalence entails a form of scale invariance: if fundamental interactions are invariant under arbitrary decompositions of $\mathbf{U}$ into $\mathbf{S}$, $\mathbf{O}$ and $\mathbf{E}$, then they are invariant under changes in the scales at which $\mathbf{S}$, $\mathbf{O}$ and $\mathbf{E}$ are defined.  Both $\mathbf{S}$ and $\mathbf{O}$, in particular, can be defined to be arbitrarily small or large, subject only to the constraint that $\mathbf{S}$, $\mathbf{O}$ and $\mathbf{E}$ together decompose $\mathbf{U}$.  Decompositional equivalence requires, therefore, that there be no preferred scale of observation and no preferred scale of observers.  If $\mathbf{U}$ satisfies decompositional equivalence, fundamental interactions in $\mathbf{U}$ are independent of what \textit{any} observer at any scale might choose to designate as the ``system of interest.''

It is important to emphasize that this principle of decompositional equivalence concerns the decomposition of \textit{systems into subsystems} and says nothing in particular about the decomposition of states into substates or component states.  Hence the claim that a composite system $\mathbf{X}$ satisfies decompositional equivalence implies nothing in particular about the state occupied by $\mathbf{X}$; in particular, it does not imply that the state occupied by $\mathbf{X}$ is separable in the quantum-theoretic sense, as discussed in \S 5 and \S 6 below.

Noting that any decomposition of a system into subsystems effectively draws boundaries, equivalent to the voxel boundaries mentioned above, around the subsystems, the principle can alternatively be stated as:
\begin{quote}
\textbf{Decompositional equivalence} (Variant 1):  All fundamental physical interactions are entirely independent of boundaries drawn to decompose a physical system into subsystems.
\end{quote}
or
\begin{quote}
\textbf{Decompositional equivalence} (Variant 2):  The boundaries separating the subsystems of a physical system have no effect on the overall behavior of the system.
\end{quote}
This last statement of the principle is closely related to the classical ``reductionist'' principle that the behavior of a system is completely determined by the behavior of its ``microscopic'' components.  The historical reductionist program of fully predicting all macroscopic system behavior from a theory of the behavior of such microscopic components plus a set of ``bridge laws'' (e.g. Nagel, 1961) has largely been abandoned and ``emergent phenomena'' are now widely regarded as unpredictable, perhaps even in principle, from the behavior of microscopic components (e.g. Butterfield, 2011).  However, such phenomena are nonetheless still regarded as \textit{implemented by} the microscopic components of the system and hence as (possibly unpredictable) \textit{outcomes} of their fundamental interactions.  This reductionist principle has been a mainstay of the physical sciences since the early 19th century development of the classical theory of atoms; indeed, it is the principle that motivates the continuing search for yet more microscopic physical components and yet more fundamental physical interactions.  Any system to which it applies must satisfy decompositional equivalence; if it did not, macroscopic ``emergent phenomena'' could result from boundary-dependent alterations of the fundamental interactions and would thus be determined not just by the behavior of the microscopic components but by the observer's choices about the placement of subsystem boundaries.

It is worth noting explicitly that decompositional equivalence is a \textit{physical} principle, not a mathematical principle.  As such, its theoretical role is both to motivate and to constrain the choice of mathematical representations of physical systems and physical dynamics.  It requires, in particular, that any correct mathematical representation of physical dynamics must be independent of the mathematical representation of subsystem decomposition and hence of subsystem boundaries.  As will be shown in \S 5 below, the conjunction of decompositional equivalence and Landauer's principle materially implies the mathematical representation provided by standard, unitary quantum theory. If both Landauer's principle and decompositional equivalence hold, unitary quantum theory must be true; if unitary quantum theory is false, one or the other of these two principles must be false.  

Conversely, both classical and quantum physics materially imply decompositional equivalence by building it into their respective representations of the space of possible physical states.  Both theories employ abstract vector spaces to represent physical states, the real phase space in classical physics and the complex Hilbert space in quantum physics.  Both spaces can be arbitrarily decomposed into subspaces using vector-space product operators, the Cartesian product $\times$ in classical physics and the tensor product $\otimes$ in quantum physics.  These operators are both associative: for any decomposition of $\mathbf{U}$ into $\mathbf{S}$, $\mathbf{E}$ and $\mathbf{O}$, the classical phase space $\mathcal{C}_{\mathbf{U}} = \mathcal{C}_{\mathbf{S}} \times \mathcal{C}_{\mathbf{E}} \times \mathcal{C}_{\mathbf{O}} = (\mathcal{C}_{\mathbf{S}} \times \mathcal{C}_{\mathbf{E}}) \times \mathcal{C}_{\mathbf{O}} = \mathcal{C}_{\mathbf{S}} \times (\mathcal{C}_{\mathbf{E}} \times \mathcal{C}_{\mathbf{O}})$ in classical physics, and the Hilbert space $\mathcal{H}_{\mathbf{U}} = \mathcal{H}_{\mathbf{S}} \otimes \mathcal{H}_{\mathbf{E}} \otimes \mathcal{H}_{\mathbf{O}} = (\mathcal{H}_{\mathbf{S}} \otimes \mathcal{H}_{\mathbf{E}}) \otimes \mathcal{H}_{\mathbf{O}} = \mathcal{H}_{\mathbf{S}} \otimes (\mathcal{H}_{\mathbf{E}} \otimes \mathcal{H}_{\mathbf{O}})$ in quantum physics.  Both classical and quantum physics, therefore, require the state space of any system to be \textit{entirely invariant} under arbitrary product decompositions.  In both theories, moreover, physical dynamics are represented by a Hamiltonian operator that is additive and therefore associative: if $\hat{H}_{\mathbf{U}}$ is the Hamiltonian representing the dynamics of $\mathbf{U}$, then in both theories $\hat{H}_{\mathbf{U}} = \hat{H}_{\mathbf{S}} + \hat{H}_{\mathbf{E}} + \hat{H}_{\mathbf{O}} + \hat{H}_{\mathbf{SE}} + \hat{H}_{\mathbf{SO}} + \hat{H}_{\mathbf{EO}} + \hat{H}_{\mathbf{SEO}}$ for any arbitrary choice of $\mathbf{S}$, $\mathbf{E}$ and $\mathbf{O}$, where  $\hat{H}_{\mathbf{S}}$, $\hat{H}_{\mathbf{E}}$ and $\hat{H}_{\mathbf{O}}$ are the respective self-interactions and the higher-order terms are between-system interactions.  Both classical and quantum physics, therefore, require the dynamics of any system to be \textit{entirely invariant} under arbitrary additive decompositions of the system's Hamiltonian.  Were decompositional equivalence shown to be false in our universe, neither state spaces nor Hamiltonians would be arbitrarily decomposable in this way and the basic mathematical formalisms of both quantum and classical physics would have to be rejected.

As important as what these formal representations require is what they \textit{do not} require.  The classical phase-space representation does not, by itself, require that position and momentum be instantaneously and hence simultaneously measurable with arbitrary precision; it does not, in other words, require the observable state of every or even any classical system to be a point in phase space.  Hence it does not require Landauer's principle to be violated; indeed doing so would render classical statistical mechanics inconsistent.  This key component of the ``classical worldview'' is instead a separate assumption that the classical phase space merely allows.  Similarly, the state-space representations of neither classical nor quantum physics require the assumption that subsystems of every or even any composite system $\mathbf{X}$ can be individually and independently manipulated, and neither requires the assumption that observational outcomes obtained from a composite system $\mathbf{X}$ can, in every or even any case, be uniquely attributed to specific, uniquely identifiable subsystems of $\mathbf{X}$.  It is shown below that these two assumptions, both of which are central components of the classical worldview, violate the conjunction of decompositional equivalence and Landauer's principle and are therefore inconsistent with the state-space formalisms of either classical or quantum physics if Landauer's principle is assumed.  Various standard approximations and idealizations commonly employed in both classical and quantum physics make these assumptions and therefore violate decompositional equivalence, Landauer's principle or both.  Alternatives to standard, unitary quantum theory that introduce a non-unitary physical ``collapse'' process (e.g. Ghirardi, Rimini and Weber, 1986; Penrose, 1996; Weinberg, 2012), in particular, violate decompositional equivalence if the collapse process is taken to occur at some, but not all possible, boundaries in the state space.  It is the ubiquitous presence of ancillary assumptions that violate either decompositional equivalence or Landauer's principle that render typical presentations of even standard unitary quantum theory paradoxical.  When both principles are explicitly and rigorously respected, nothing collapses, nothing decoheres and there are no ``multiple worlds.''  As discussed below, a strict adherence to decompositional equivalence and Landauer's principle replaces these interpretative standbys with a deep and unresolvable ambiguity about the physical sources of observational outcomes.  The paradoxes associated with ``collapse'' are replaced, in this case, by deep empirical questions about the pragmatic abilities of observers, human or otherwise, to identify particular physical systems as sources of observational outcomes, both under ordinary circumstances and in the laboratory.

Demonstrating that decompositional equivalence and Landauer's principle together materially imply standard, unitary quantum theory is surprisingly straightforward.  The first step is to recognize that if $\mathbf{O}$ can be chosen arbitrarily, $\mathbf{O}$ cannot be assumed to encode any particular information about $\mathbf{S}$ prior to making observations.  Therefore the only information about $\mathbf{S}$ that $\mathbf{O}$ can be regarded as encoding is information that $\mathbf{O}$ has obtained by observation.  Given Landauer's principle, obtaining classically-encodable information by observation requires physical interaction.  Hence $\mathbf{O}$ can obtain information about $\mathbf{S}$ only through physical interaction.  The second step is to note that the subsystem of $\mathbf{U}$ with which $\mathbf{O}$ interacts, which will be referred to as $\mathbf{B_{O}}$ for $\mathbf{O}$'s ``box,'' comprises both $\mathbf{S}$ and $\mathbf{E}$.  Because $\mathbf{U}$ satisfies decompositional equivalence, $\mathbf{O}$'s physical interaction with $\mathbf{B_{O}}$ must be invariant under arbitrary alternative partitionings of $\mathbf{B_{O}}$, i.e. arbitrary alternative placements of the $\mathbf{S}$ - $\mathbf{E}$ boundary.  This invariance of the $\mathbf{O}$ - $\mathbf{B_{O}}$ interaction under arbitrary alternative partitionings of $\mathbf{B_{O}}$ renders $\mathbf{B_{O}}$ a black box: if the observable behavior of $\mathbf{B_{O}}$ is invariant under arbitrary rearrangements of any subsystem boundaries within $\mathbf{B_{O}}$, then $\mathbf{O}$'s observations of this behavior \textit{a fortiori} can yield no information about such subsystem boundaries and hence no information about the ``internal structure'' or ``internal dynamics'' of $\mathbf{B_{O}}$.  The third step is to show that if Landauer's Principle is respected, the acquisition of information from a black box must be described by unitary quantum theory.  The three sections that follow make this demonstration precise.

\section{Information can be obtained only by observation}

As before, let $\mathbf{S}$, $\mathbf{E}$ and $\mathbf{O}$ be physical subsystems that jointly decompose the universe $\mathbf{U}$.  If $\mathbf{U}$ satisfies decompositional equivalence, $\mathbf{S}$, $\mathbf{E}$ and $\mathbf{O}$ can be defined or designated arbitrarily; the labels `$\mathbf{S}$,' `$\mathbf{E}$' and `$\mathbf{O}$' can, therefore, have no special meaning.  Any subsystem whatsoever can be designated as the system of interest; similarly, any subsystem whatsoever can be considered to be an ``environment'' or an ``observer.''  For a fixed $\mathbf{S}$, in particular, the $\mathbf{E}$ - $\mathbf{O}$ boundary can be varied arbitrarily.  Any restriction on the kinds of subsystems labelled with these terms, e.g. any restriction on the Hamiltonians $\hat{H}_{\mathbf{S}}$, $\hat{H}_{\mathbf{E}}$ or $\hat{H}_{\mathbf{O}}$ beyond the requirement that $\hat{H}_{\mathbf{U}} = \hat{H}_{\mathbf{S}} + \hat{H}_{\mathbf{E}} + \hat{H}_{\mathbf{O}} + \hat{H}_{\mathbf{SE}} + \hat{H}_{\mathbf{SO}} + \hat{H}_{\mathbf{EO}} + \hat{H}_{\mathbf{SEO}}$, violates decompositional equivalence.

The statement that $\mathbf{O}$, the ``observer'' can be chosen arbitrarily naturally raises the question of what ``observation'' means.  Intuitively, to make an observation is to obtain a particular, determinate observational outcome; to make an observation is, in other words, to obtain \textit{classical} information, information that can be encoded by a finite string of bits.  Landauer's principle requires classical information to be encoded in a thermodynamically-irreversible way, e.g. recorded on some physical memory.  Hence an ``observer'' must have at least one physical degree of freedom that can change in a thermodynamically-irreversible way.  Setting aside the question of how thermodynamic irreversibility is physically implemented until \S 5 below, this requirement has two implications.  First, $\mathbf{O}$ cannot be an ``empty'' subsystem of $\mathbf{U}$; it must contain at least one physical degree of freedom.  An observer cannot, in other words, be a merely notional coordinate system as is sometimes assumed.  This has the important consequence that \textit{all observers are inside} $\mathbf{U}$ and hence that $\mathbf{U}$ \textit{cannot be observed} even in principle.  There is, therefore, no ``observable state of $\mathbf{U}$'' even in principle.  The second consequence is that the physical state of $\mathbf{O}$ must be well-defined \textit{in the context of the} $\mathbf{S}$ - $\mathbf{E}$ - $\mathbf{O}$ \textit{decomposition of} $\mathbf{U}$.  Note that this does not require that the degree(s) of freedom composing $\mathbf{O}$ must have a well-defined state in the context of any other decomposition of $\mathbf{U}$, and does not require that either $\mathbf{S}$ or $\mathbf{E}$ have a well-defined state in any decomposition.  Hence the separability, in the quantum-theoretic sense, of the unobservable ``state'' of $\mathbf{U}$ is not required by the designation of some subsystem of $\mathbf{U}$ as an observer.

While it appears trivial from a formal perspective, the arbitrariness with which $\mathbf{S}$, $\mathbf{E}$ and $\mathbf{O}$ can be chosen has important consequences for how any theory consistent with decompositional equivalence is interpreted and used.  The first and most important of these is that observers cannot be assumed to have any special characteristics or to be in any particular special state at the initiation of observations.  Observers cannot, in particular, be assumed to have any prior knowledge of the system being observed.  If observation is characterized in Bayesian terms, the probability distribution over the states of the observed system prior to any observations being conducted must be assumed to be uniform.

The assumption that observers have no prior knowledge of the system being observed is often stated in discussions of both classical and quantum physics.  Schlosshauer, for example, includes the idea that observers can be ``initially completely ignorant'' when describing classical observations:
\begin{quote}
``Here (i.e. in classical physics) we can enlarge our `catalog' of physical properties of the system (and therefore specify its state more completely) by performing an arbitrary number of measurements of identical physical quantities, in any given order.  Moreover, many independent observers may carry out such measurements (and agree on the results) without running into any risk of disturbing the state of the system, even though they may have been initially completely ignorant of this state.''
\begin{flushright}
Schlosshauer, 2007, p. 16
\end{flushright} 
\end{quote}
Ollivier, Poulin and Zurek employ similar language to operationally define \textit{objectivity} in a quantum-theoretic context:
\begin{quote}
``A property of a physical system is \textit{objective} when it is:
\begin{list}{\leftmargin=2em}
\item
1. simultaneously accessible to many observers,
\item
2. who are able to find out what it is without prior knowledge about the system of interest, and 
\item
3. who can arrive at a consensus about it without prior agreement.''
\end{list}
\begin{flushright}
Ollivier, Poulin and Zurek, 2004, p. 1; Ollivier, Poulin and Zurek, 2005, p. 3
\end{flushright}
\end{quote}
Both of these statements, however, leave unmentioned and unaddressed a fundamental question: how do the many independent observers \textit{identify} the system being jointly observed?  If they are \textit{completely ignorant} of the system, having \textit{no prior knowledge} about it, then they cannot know, for example, where it is located, what it looks like, its mass, or its velocity relative to their own.  How then can they agree that they have observed the same system?  Identifying the system of interest, it would seem, \textit{requires} observers to have \textit{a priori} knowledge of the system to be identified.

The above two passages illustrate two common and often implicit assumptions: 1) that the observations employed to identify the system of interest are unproblematic and do not need to be considered as part of a theoretical description of the observation process, and 2) that multiple observers can, by observation, identify \textit{exactly the same} physical system.  They also illustrate a common, subtle but important equivocation in the use of the term ``system.''  When ``system'' refers to what the observers jointly identify, it refers to something, such as an item of apparatus, about which the observers are assumed to have considerable \textit{a priori} information.  The observers are, in particular, assumed to know where the apparatus is, what it looks like, and so forth.  When, on the other hand, ``system'' refers to what the observers do not have any \textit{a priori} knowledge about, it refers \textit{only} to the collection of degrees of freedom that are being probed by the experiment.  This latter usage has been emphasized explicitly by Tegmark (2000, 2012) in his discussions of decoherence.  As all components of the state of the ``system'' in this latter usage are unknown by definition, this latter system cannot, even in principle, be identified by observation.  Hence the above passages also reveal a third common assumption: that identifying the apparatus, or more precisely, the prepared state of the apparatus, is sufficient to render \textit{all and only} the degrees of freedom that the experiment is designed to probe observable.  This assumption can, informally, be stated as the assumption that ``preparing the apparatus is preparing the system'' where ``system'' is here used in the second of the above two senses.

As will be shown below, the three assumptions made explicit above all violate decompositional equivalence.  Formal descriptions of measurement that stipulate, either explicitly from a theoretical or ``god's eye'' (Koenderink, 2014) perspective or implicitly through the use of notation, that multiple observers have identified and are observing exactly the same system not only beg the question of how this could have been accomplished, but are inconsistent with the quantum formalism, which as noted above enforces decompositional equivalence through both its state space description and the additivity of the Hamiltonian.  If the multiple observers can be arbitrarily chosen and cannot be assumed to have any prior knowledge of the system to be observed, all that can be said is that each observer interacted with a system -- \textit{some system or other} -- and received one or more observational outcomes.  The observational outcome(s) received constitute each observer's entire knowledge of the state of the system(s) with which they interacted.   As noted in \S 8 below, the question of \textit{how} multiple observers can, in practice, arrive at an agreement that they have observed the same, or at least equivalent systems becomes an empirical question that physics must answer.

\section{Observation is interaction with a black box}

A well-established, over half-century-old body of theory describes the situation in which an observer can obtain outcomes from an observed system, but can know nothing about the system beyond the outcomes that have been obtained.  This is precisely the situation of an observer who interacts with a ``black box'' as described by classical cybernetics (e.g. Ashby, 1956; Moore, 1956) or of an observer who receives signals, via a communication channel, from an unknown source as described by classical information theory (Shannon, 1948).  A black box is, by definition, a system with observable overt behavior but an inaccessible, unobservable and hence unknowable interior.  Ashby (1956) and Moore (1956) independently proved that, while observations of the overt behavior of a black box can clearly place a \textit{lower} limit on the complexity of the unknown mechanism inside the box, no finite set of observations of the box's behavior can place an \textit{upper} limit on its internal complexity.  The very next behavior of any black box can be a complete surprise, one that indicates the existence of unanticipated internal degrees of freedom and unanticipated internal dynamics.

It is important to note that the term ``black box'' implies an \textit{epistemic} boundary between the observer and the system observed, not a spatial separation.  Hence one can, for example, characterize a stock market as a ``black box'' for any observer who has access only to the past and current stock prices.  It remains a black box even for an observer standing in the middle of its trading floor.  The global climate is similarly a black box, even for those of us embedded in it.  Indeed Ashby viewed the theory of the black box as universal: ``The theory of the Black Box is merely the theory of real objects or systems, when close attention is given to the question, relating object and observer, about what information comes from the object, and how it is obtained'' (Ashby, 1956, p. 110).

Consider now the behavior of a universe $\mathbf{U}$ that has been arbitrarily partitioned into two subsystems $\mathbf{O}$ and $\mathbf{B_{O}}$ as shown in Fig. 1.  Decompositional equivalence requires that this partitioning has no effect on the dynamics of $\mathbf{U}$ and indeed, no physical consequences whatsoever; in particular, \textit{no physical separation} between $\mathbf{O}$ and $\mathbf{B_{O}}$ is implied.  Let us assume, for the purposes of argument, that the self-interaction of $\mathbf{U}$ can be represented by an operator (notationally anticipating a Hamiltonian) $\hat{H}_{\mathbf{U}}$, making no assumption about the structure of this operator other than that it is additively decomposable. The partition between $\mathbf{O}$ and $\mathbf{B_{O}}$ then allows this interaction operator to be written as $\hat{H}_{\mathbf{U}} = \hat{\mathnormal{H}}_{\mathbf{O}} + \hat{\mathnormal{H}}_{\mathbf{B_{O}}} + \hat{\mathnormal{H}}_{\mathbf{OB_{O}}}$, where the last term represents the interaction between the partitions.

\begin{figure}
\centering
\includegraphics[width=160mm]{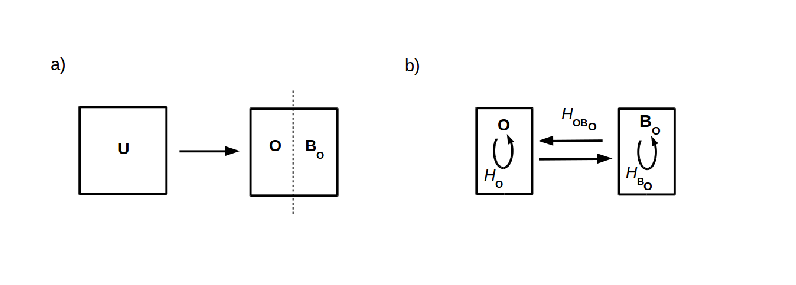}
\caption{a) A system $\mathbf{U}$ is partitioned into two subsystems $\mathbf{O}$ and $\mathbf{B_{O}}$.  b) The systems $\mathbf{O}$ and $\mathbf{B_{O}}$ have internal dynamics $\hat{H}_{\mathbf{O}}$ and $\hat{H}_{\mathbf{B_{O}}}$ respectively and interact via $\hat{H}_{\mathbf{OB_{O}}}$.}
\end{figure}

In this case, the interaction $\hat{H}_{\mathbf{OB_{O}}}$ can be regarded as defining a classical information channel between $\mathbf{O}$ and $\mathbf{B_{O}}$ and $\mathbf{O}$ can be regarded as obtaining observational outcomes by interacting with $\mathbf{B_{O}}$.  The channel defined by $\hat{H}_{\mathbf{OB_{O}}}$ is, moreover, the \textit{only} information channel between $\mathbf{O}$ and $\mathbf{B_{O}}$.  In particular, $\mathbf{O}$ has no independent access to the internal dynamics $\hat{H}_{\mathbf{B_{O}}}$ of $\mathbf{B_{O}}$.  The system $\mathbf{B_{O}}$ is, under these conditions, a black box from the perspective of $\mathbf{O}$; hence the nomenclature, ``$\mathbf{O}$'s box'' for $\mathbf{B_{O}}$ introduced earlier.

Note that the labels `$\mathbf{O}$' and `$\mathbf{B_{O}}$' can be re-interpreted so as to designate $\mathbf{B_{O}}$ the ``observer'' and $\mathbf{O}$ the ``observed system.''  In this case $\mathbf{B_{O}}$ obtains outcomes from $\mathbf{O}$ via the information channel defined by $\hat{H}_{\mathbf{OB_{O}}}$; from the perspective of $\mathbf{B_{O}}$, $\mathbf{O}$ is a black box.

The system $\mathbf{B_{O}}$ being a black box from $\mathbf{O}$'s perspective has an immediate consequence of interest for any physical theory constructed by $\mathbf{O}$: since any further partition of $\mathbf{B_{O}}$ is internal to $\mathbf{B_{O}}$, $\mathbf{O}$ can obtain no information about any such internal partition.  This becomes obvious when the relevant interactions are examined.  If $\mathbf{B_{O}}$ is further partitioned into $\mathbf{S}$ plus $\mathbf{E}$, the internal interaction $\hat{H}_{\mathbf{B_{O}}}$ can be written as $\hat{H}_{\mathbf{B_{O}}} = \hat{\mathnormal{H}}_{\mathbf{S}} + \hat{\mathnormal{H}}_{\mathbf{E}} + \hat{\mathnormal{H}}_{\mathbf{SE}}$.  As $\mathbf{O}$ has no access to $\hat{H}_{\mathbf{B_{O}}}$, $\mathbf{O}$ can have no access to the internal self-interactions $\hat{H}_{\mathbf{S}}$ and $\hat{H}_{\mathbf{E}}$ or to the two-way interaction $\hat{H}_{\mathbf{SE}}$.  Indeed decompositional equivalence guarantees that the interaction $\hat{H}_{\mathbf{OB_{O}}}$ via which $\mathbf{O}$ acquires observational outcomes from $\mathbf{B_{O}}$ is entirely independent of arbitrary redefinitions of the $\mathbf{S}$-$\mathbf{E}$ partition and hence arbitrary redefinitions of the interaction $\hat{H}_{\mathbf{SE}}$.  Hence in any universe satisfying decompositional equivalence, \textit{observers cannot ``see'' system-environment boundaries} and \textit{observers cannot observationally characterize system-environment interactions}.  The split between ``system'' and ``environment'' is both \textit{merely notional} and \textit{entirely arbitrary}; it has no effect whatsoever on the observable behavior of $\mathbf{B_{O}}$.  Observers cannot, in particular, observationally characterize individual states of either system or environment.  All observers can do is obtain outcomes, via $\hat{H}_{\mathbf{OB_{O}}}$, from $\mathbf{B_{O}}$, the black box within which the subsystems $\mathbf{S}$ and $\mathbf{E}$, however they or their interaction may be defined, are fully contained.

This limitation on the information actually obtainable by observers immediately implies that the three common assumptions about observability made explicit in the last section are illegitimate in any theory respecting both decompositional equivalence and Landauer's principle.  Observers cannot, in particular, be assumed to identify systems unproblematically; indeed as discussed in \S 8 below, how observers manage to identify systems in practice becomes a compelling empirical question.  Multiple observers cannot be assumed to identify exactly the same system, or even to uniformly distinguish each other from the surrounding shared environment.  Finally, even the identification of all of $\mathbf{B_{O}}$ as a prepared apparatus cannot be assumed to entail any specific consequences for any unobserved degrees of freedom within $\mathbf{B_{O}}$; as with any black box, the very next behavior of $\mathbf{B_{O}}$ can be a complete surprise that reveals the existence of unanticipated and uncharacterized degrees of freedom.  It should be emphasized, moreover, that no specifically quantum-mechanical assumptions have been made in reaching these conclusions; they follow solely from \textit{classical} cybernetics or information theory when decompositional equivalence and Landauer's principle are respected.  

The results of this section also show that the assumptions made in standard environmental-decoherence calculations (e.g. Schlosshauer, 2007; Zurek, 2003) cannot be given anything other than a pragmatic, \textit{post-hoc} justification.  Observers cannot, in particular, either identify the system-environment boundary at which decoherence is taken to occur or establish that the environmental state is sufficiently random that system-environment entanglement results in decoherence.  While \textit{post-hoc} justifications of the assumptions required for decoherence calculations may be adequate in practical settings, they cannot support in-principle foundational claims; in particular, they cannot support the common claim that decoherence \textit{explains} the ``emergence of classicality'' from unitary dynamics (Fields, 2011; Fields, 2014a; Kastner, 2014).  To claim that decoherence explains the emergence of classical system boundaries when such boundaries must be assumed to perform decoherence calculations is to beg the question.  To claim that decoherence explains the emergence of classical system states (``pointer states'') when an effectively classical state of the environment must be assumed to perform the calculation is also to beg the question.

\section{Decompositional equivalence and Landauer's principle together imply quantum theory}

\subsection{Overview}

The primary claim of this paper is that decompositional equivalence and Landauer's Principle, taken together, materially imply standard, unitary quantum theory.  Quantum theory can be false, therefore, only if one or the other of these two principles is violated.  Its secondary claim is that when quantum theory is seen as an inevitable result of decompositional equivalence plus Landauer's Principle, it ceases to be paradoxical.  In particular, the need for a paradoxical ``collapse'' process or for ``multiple worlds'' disappears.  Indeed, when it is seen as a consequence of these two principles, quantum theory appears to be simple and intuitively compelling.  It appears, in other words, as a theory we should \textit{expect} to be true from first principles.

To show that quantum theory follows from decompositional equivalence plus Landauer's principle, we consider a thought experiment that illustrates the simplest informative interaction between an observer and the world: one that yields a one-bit outcome.  By working through this thought experiment, we show that action must be quantized and that the principal axioms (their statements in Zurek (2003) are taken to be canonical; similar statements can be found in Nielsen and Chaung (2000) and elsewhere) are implied by decompositional equivalence plus Landauer's principle.  In each case, the formal representation required by the axiom is first shown to arise naturally under the conditions imposed by the thought experiment.  The contrapositive statement, that the axiom cannot be false without violating either decompositional equivalence or Landauer's principle, is then demonstrated.  In so doing, we show why a number of common assumptions cannot be made without violating either decompositional equivalence or Landauer's principle and hence contradicting quantum theory.  This sets the stage for an intuitive understanding, in \S 6, of why quantum theory has the structure that it does.  

\subsection{Thermodynamics of observation}

Consider an observer $\mathbf{O}$ capable of receiving and recording only a one-bit outcome.  Suppose, for example, that $\mathbf{O}$ is equipped with a horizontally-oriented meter stick that has been modified to produce a digital ``1'' signal if but only if it comes into contact with an object that is 1 m wide and to produce a ``0'' signal if it comes into contact with any object that is not 1 m wide.  Suppose, moreover, that this observer is embedded in a world $\mathbf{B_{O}}$ containing many different objects, some number $n \gg 1$ of which have a horizontal dimension of 1 m while $m \gg 1$ others have other horizontal dimensions.  For simplicity, we can treat $\mathbf{O}$ as fixed and unmoving, and imagine that some mechanism external to $\mathbf{O}$ occasionally moves an object into contact with $\mathbf{O}$'s meter stick.  As before, $\mathbf{B_{O}}$ is taken to include everything in the universe that is not part of $\mathbf{O}$.  Figure 2 illustrates this situation, indicating by the observer's blindfold that \textit{only} the ``1'' and ``0'' outcomes generated by the meter stick can be detected by $\mathbf{O}$.  From $\mathbf{O}$'s perspective, $\mathbf{B_{O}}$ is clearly a black box: in particular, $\mathbf{O}$ has no observational access to the mechanism that occasionally brings objects into contact with the meter stick, no means of counting the objects, and no means other than the meter stick of distinguishing them.  This situation clearly generalizes to any finite number of measurement devices yielding any finite number of finite-resolution outcomes.  It represents, therefore, the situation faced, in practice, by all observers that are restricted to expending finite energy and hence to recording finite observational outcomes.

\begin{figure}
\centering
\includegraphics[width=160mm]{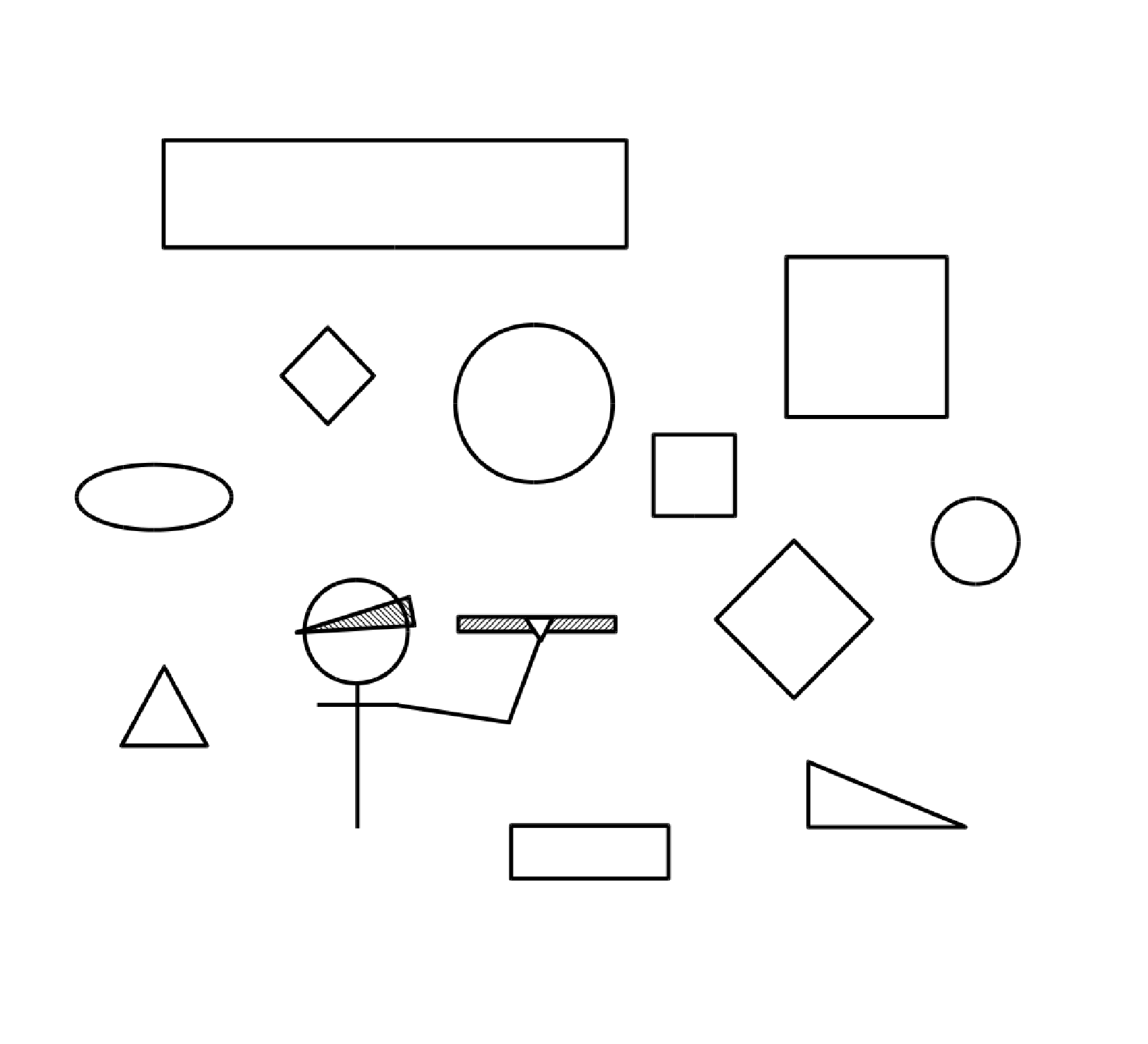}
\caption{A blindfolded observer equipped only with a horizontally-oriented meter stick, embedded in a world containing multiple objects, some but not all of which have a horizontal dimension of 1 m.}
\end{figure}

As $\mathbf{O}$ explores $\mathbf{B_{O}}$, the meter stick occasionally produces a ``1'' signal indicating the presence of a 1 m wide object; it also occasionally produces a ``0'' signal indicating the presence of something else.  Let us now invoke Landauer's principle: recording each of these outcomes requires a thermodynamically-irreversible change in the physical state of $\mathbf{O}$.  Each of these outcome-recording state changes must, therefore, consume at least $0.7 kT_{\mathbf{O}}$ of free energy, where $k$ is Boltzmann's constant and $T_{\mathbf{O}} > 0$ is the temperature of $\mathbf{O}$; Anders \textit{et al.} (2006) show that this is the case even for strongly-coupled systems, i.e. even if the local interaction is far from equilibrium.  For simplicity, let us assume maximal efficiency, so that each state change requires exactly $0.7 kT_{\mathbf{O}}$.   This free energy must be supplied either by $\mathbf{O}$ itself, e.g. from an on-board battery that powers the generation of signals by the digital meter stick, or supplied by $\mathbf{B_{O}}$, e.g. by extracting energy from each object the meter stick contacts.  In the first case, each bit recorded by $\mathbf{O}$ dissipates $0.7 kT_{\mathbf{O}}$ into $\mathbf{B_{O}}$; in the second, each bit recorded extracts $0.7 kT_{\mathbf{O}}$ from $\mathbf{B_{O}}$.  The $\mathbf{O}$ - $\mathbf{B_{O}}$ interaction, which anticipating its characterization as a Hamiltonian will be written $\hat{H}_{\mathbf{OB_{O}}}$, can, therefore, be considered an energy observable from either the perspective of $\mathbf{O}$ or the perspective of $\mathbf{B_{O}}$; $\mathbf{O}$ can, in particular, be regarded as ``counting'' energy increments of $0.7 kT_{\mathbf{O}}$.  There is, however, no ``external'' perspective from which the energy states of both systems can be simultaneously determined; $\mathbf{O}$ and $\mathbf{B_{O}}$ together constitute the whole universe $\mathbf{U}$, and there is nothing outside of $\mathbf{U}$ that can observe its state.

Let us now further simplify the situation by assuming that these outcome-recording state changes are the \textit{only} state changes that $\mathbf{O}$ undergoes.  In this case, $\mathbf{O}$'s state changes can be viewed as implementing a \textit{clock} that defines an observer-relative time coordinate $t_{\mathbf{O}}$.  Let $\Delta t_{\mathbf{O}}$ be the period of this clock.  Since $\mathbf{O}$'s state changes \textit{only} when this clock ``ticks,'' the period $\Delta t_{\mathbf{O}}$ can be defined as constant without loss of generality.

It is with respect to this internally-defined, observer-relative time coordinate $t_{\mathbf{O}}$ that $\mathbf{O}$'s state changes can be defined as ``irreversible'' and hence ``classical''; indeed $\mathbf{O}$'s state changes are irreversible precisely \textit{because} they can be viewed as implementing a clock.  From $\mathbf{O}$'s ``internal'' perspective, this corresponds to their being both sequential and distinct from each other, both of which are intuitive characteristics of classicality.  Note that in this situation, classicality has not ``emerged'' from the physical dynamics of the $\mathbf{O}$ - $\mathbf{B_{O}}$ interaction.  There is nothing about the dynamics that \textit{imposes} the interpretation of $\mathbf{O}$'s state changes as counting increments of transferred energy.  This is rather a convenient theoretical description of the dynamics from the observer's own, or from a theorist's, perspective that associates bits as units of information transfer with $0.7 kT_{\mathbf{O}}$ increments of energy transfer and hence defines time increments as an alternative notation for energy increments. 

This intuitive connection between incremented time and classicality can perhaps be sharpened by contrasting classical and quantum descriptions of a particle in a one-dimensional box.  In the classical case, each collision of the particle with the wall of the box results in a reversal of the momentum of the particle; it can therefore be viewed as a measurement by the particle of the position of the wall.  Each such measurement dissipates an increment of kinetic energy from the particle into the wall.  The back-and-forth bounces of the particle can, therefore, be viewed as implementing a clock; as the particle loses energy to the wall, the clock slows down and finally stops.  In the quantum case, on the other hand, there are no discrete events in which the momentum reverses; the particle can, therefore, be represented as a standing wave.  In this case, nothing is measured, no energy is lost and from the particle's perspective, there is no elapsed time.  The $t$ coordinate in the Schr\"{o}dinger equation is, in this case, not an observable and indeed has no physical meaning.

\subsection{Quantization of action}

As noted above, the interaction $\hat{H}_{\mathbf{OB_{O}}}$ transfers $0.7 ~kT_{\mathbf{O}}$ between $\mathbf{O}$ and $\mathbf{B_{O}}$ during the time increment $\Delta t_{\mathbf{O}}$.  The minimum action of $\hat{H}_{\mathbf{OB_{O}}}$ is then:

\begin{equation}
0.7 ~kT_{\mathbf{O}}\Delta t_{\mathbf{O}} = \int_{\Delta t_{\mathbf{O}}} \hat{\mathnormal{H}}_{\mathbf{OB_{O}}} \mathnormal{dt_{\mathbf{O}}}.
\end{equation}
As decompositional equivalence allows $\mathbf{O}$ to be chosen arbitrarily, let us assume that $\mathbf{O}$ has been chosen in such a way that $0.7 ~kT_{\mathbf{O}}\Delta t_{\mathbf{O}}$ is minimal across all two-way partitions of $\mathbf{U}$ into an observer and that observer's world.  We can then define $h^{\prime} = 0.7 ~kT_{\mathbf{O}}\Delta t_{\mathbf{O}}$ as the \textit{minimal action to receive and encode 1 bit} in $\mathbf{U}$ and call it a ``quantum'' of action.  

Taking $\mathbf{U}$ to be our universe and supposing that the energy efficiency of biological photoreceptors has been optimized by evolutionary processes, it seems reasonable to estimate a numerical value of $h^{\prime}$ by considering such systems.  For a molecule $\mathbf{m}$ of rhodopsin at $T_{\mathbf{m}} = 310$ K (i.e. 37 C, physiological temperature), $\Delta t_{\mathbf{m}} \sim 200$ fs (Wang \textit{et al.}, 1994); other biological photoreceptors have similar response times.  In this case $kT_{\mathbf{m}} \sim 4.3 \cdot 10^{-21}$ J and the 1-bit information transfer action is $0.7 ~kT_{\mathbf{m}}\Delta t_{\mathbf{m}} \sim 6.0 \cdot 10^{-34}$ J$\cdot$s, a value remarkably close to that of Planck's constant $h \sim 6.6 \cdot 10^{-34}$ J$\cdot$s.  We can, therefore, identify $h^{\prime}$ as $h$.  

Note that this definition of $h$ is entirely thermodynamic, uses no particularly ``quantum'' concepts and does not appeal to any distinction between ``microscopic'' and ``macroscopic'' objects or behaviors.  Note also that what is ``quantized'' by this definition of $h$ is the action of \textit{transferring one bit of classical information}; indeed if no bits are transferred, there is no elapsed time and hence no action.  Jennings and Leifer (2016) have emphasized in a recent review that a quantum of action can easily be reproduced within a purely-classical model of measurement.  One might speculate that had classical information theory been developed before quantum theory, rather than the reverse, not just the quantization of action but much else about the theory might not have been viewed as physically paradoxical.

\subsection{Observationally-distinguishable states}

Let us now adopt an explicitly theoretical, ``god's eye'' perspective on the $\mathbf{O}$ - $\mathbf{B_{O}}$ interaction; indeed this perspective is already implicit in Fig. 2.  From this perspective, the ``ontic'' physical state of $\mathbf{B_{O}}$ can be described as a spatial configuration of objects, some one of which may, but need not, be in contact with $\mathbf{O}$'s meter stick.  The mechanism that occasionally moves an object into contact with $\mathbf{O}$'s meter stick can be considered to be implemented by a self-interaction $\hat{H}_{\mathbf{B_{O}}}$ (again notationally anticipating the Hamiltonian) that acts on these states.  As this is the only mechanism that affects the state of $\mathbf{O}$, we can assume for simplicity that this is the only mechanism implemented by $\hat{H}_{\mathbf{B_{O}}}$.  Both the states and $\hat{H}_{\mathbf{B_{O}}}$ can, from this perspective, be considered classical as further discussed in \S 5.7 below. 

The ``god's eye'' perspective is not, however, the perspective of any observer; no measurements can be made from this perspective.  What is of interest is how the observable results of the action of the dynamics $\hat{H}_{\mathbf{B_{O}}}$ can be described from $\mathbf{O}$'s perspective, using only the information obtainable by observation to construct the description.  This limited, observer's perspective, not the theoretical, ``god's eye'' perspective, is the one that we, as human observers, in fact experience.  A theory inferred from this perspective is the only kind of theory that can be constructed on the basis of observational evidence.  As we will see, a \textit{physical} theory inferred from this perspective, i.e. a theory that describes transitions between \textit{observationally-distinguishable states}, has the formal structure of quantum theory.

As the only observational outcomes that $\mathbf{O}$ can obtain from $\mathbf{B_{O}}$ are the values ``1'' and ``0,'' $\mathbf{O}$ can attribute at most two observationally-distinguishable states to $\mathbf{B_{O}}$.  Let us call these two states $| 1 \rangle$ and $| 0 \rangle$.  These states are clearly ``observer relative'' in the sense introduced by Rovelli (1996); they are defined solely on the basis of the observational outcomes that $\mathbf{O}$ can detect.   They are also ``relational'' in the deeper sense of being states imputed to $\mathbf{B_{O}}$ by $\mathbf{O}$, where both $\mathbf{O}$ and $\mathbf{B_{O}}$ are themselves only defined relative to the $\mathbf{O}$ - $\mathbf{B_{O}}$ decomposition.  By determining the interaction $\hat{H}_{\mathbf{OB_{O}}}$ and thereby determining the outcomes that $\mathbf{O}$ can obtain, the $\mathbf{O}$ - $\mathbf{B_{O}}$ decomposition fully determines the observationally-distinguishable states of $\mathbf{B_{O}}$.  Decomposition itself, therefore, resolves the ``choice of basis problem'' (e.g. Zurek, 2003).  It is only when the decomposition is underspecified -- typically by underspecifying the structure and hence the measurement capabilities of and outcomes obtainable by $\mathbf{O}$ -- that the interaction Hamiltonian and hence the interaction basis appears to require a decomposition-independent specification.  Once the outcomes obtainable by $\mathbf{O}$ are fixed by a full specification of $\hat{H}_{\mathbf{OB_{O}}}$, any uncertainty about the measurement basis vanishes.

From $\mathbf{O}$'s perspective, then, a sequence of observed outcomes $\dots$ 1, 1, 0, 1, 0, 0, 0, $\dots$ corresponds \textit{by definition} to a sequence of observationally-distinguishable states $\dots~ | 1 \rangle$, $| 1 \rangle$, $| 0 \rangle$, $| 1 \rangle$, $| 0 \rangle$, $| 0 \rangle$, $| 0 \rangle ~\dots$ of $\mathbf{B_{O}}$.  The effect on these states of the self-interaction $\hat{H}_{\mathbf{B_{O}}}$ can be represented by a two-component function $\{ \hat{E}_{i} \} = \{ \hat{E}_{0}, \hat{E}_{1} \}$ where:

\begin{equation}
\hat{E}_{0}: \mathrm{| 0 \rangle}, \mathrm{| 1 \rangle} \mapsto \mathrm{| 0 \rangle}; 
~\hat{\mathnormal{E}}_{1}: \mathrm{| 0 \rangle}, \mathrm{| 1 \rangle} \mapsto \mathrm{| 1 \rangle}.
\end{equation}
The above sequence of outcomes then corresponds to the sequence $\dots~ \hat{E}_{1}, \hat{E}_{1}, \hat{E}_{0}, \hat{E}_{1}, \hat{E}_{0}, \hat{E}_{0}, \hat{E}_{0}, ~\dots$ of operator actions.  These two components $\hat{E}_{0}$ and $\hat{E}_{1}$ of $\{ E_{i} \}$ are clearly orthogonal.  Letting $\mathcal{H}_{\mathbf{B_{O}}}$ be an abstract space of to-be-characterized structure that contains $| 0 \rangle$ and $| 1 \rangle$ -- we will see below that $| 0 \rangle$ and $| 1 \rangle$ are in fact basis vectors of $\mathcal{H}_{\mathbf{B_{O}}}$ -- the components $\hat{E}_{0}$ and $\hat{E}_{1}$ clearly resolve the identity on $\mathcal{H}_{\mathbf{B_{O}}}$.

On the physically-reasonable assumption that the abstract state space $\mathcal{H}_{\mathbf{B_{O}}}$ is measurable, the function $\{ E_{i} \}$ is a positive operator-valued measure (POVM) on $\mathcal{H}_{\mathbf{B_{O}}}$, as required by the measurement axiom of quantum theory (axiom iii in Zurek, 2003).  Hence the requirement that measurements be representable by POVMs is a consequence of just two intuitively quite natural ideas: that any operation on any space only makes sense as a representation of measurement if the space is measurable, and that the set of outcomes obtainable by any observer is well-defined and, if Landauer's principle is to be respected, finite (that observations can have only finite numbers of recordable outcomes has been stressed previously, e.g. in Peres and Terno, 2004).  A space fails to be measurable, however, only if it has no $\sigma$-algebra of subsets over which a real-valued function that commutes with subset union can be defined (e.g. Jauch, 1968).  Decompositional equivalence requires that subsystem boundaries have no effect on how a system works; it thus requires \textit{any} function that characterizes a system's overall behavior to commute with subsystem union.  A space that fails to be measurable cannot, therefore, represent a system that satisfies decompositional equivalence.  Hence if measurements of a system's behavior cannot be represented by POVMs, the system cannot satisfy decompositional equivalence.  Decompositional equivalence materially implies, therefore, the representability of measurement by POVMs.  The POVM representation and the labelling of observationally distinguishable states by outcome values together immediately imply that an observation $o$ leaves the system in the observationally distinguishable state $|o\rangle$ (axiom iv in Zurek, 2003) as indeed shown already in Zurek (2003).  As all such states are observer-relative as noted above, no \textit{physical} ``collapse'' is implied by this notation; indeed any boundary-dependent physical collapse would violate decompositional equivalence.

\subsection{Unitarity}

From the theorist's perspective illustrated in Fig. 2, it is clear that the state designations ``$| 0 \rangle$'' and ``$| 1 \rangle$'' are highly ambiguous; both refer to multiple spatial configurations of objects that are distinguishable from a ``god's eye'' perspective but are indistinguishable by $\mathbf{O}$.  The state transitions represented by $\hat{E}_{0}$ and $\hat{E}_{1}$ are similarly, from the theorist's perspective, implemented by multiple distinct changes in the spatial configuration of objects.  These operators, therefore, under-represent the unknown mechanism that brings objects into contact with $\mathbf{O}$'s meter stick.  Note that this is true even if there are, in fact, only two distinct objects in $\mathbf{B_{O}}$, as nothing in the specification of $\hat{E}_{0}$ and $\hat{E}_{1}$ can capture this fact.  Let us represent this unknown mechanism, from $\mathbf{O}$'s perspective, by an operator $\hat{U}_{\mathbf{B_{O}}}: \mathcal{H}_{\mathbf{B_{O}}} \rightarrow \mathcal{H}_{\mathbf{B_{O}}}$ acting on the state space $\mathcal{H}_{\mathbf{B_{O}}}$, making at this point no assumption about the structure of $\mathcal{H}_{\mathbf{B_{O}}}$ other than that it is measurable.

As $\mathbf{O}$ functions as a clock, the time coordinate $t_{\mathbf{O}}$ can be used to parameterize the dynamics that occur in $\mathbf{B_{O}}$ and hence to parameterize the action of the unknown $\hat{U}_{\mathbf{B_{O}}}$.  Each tick of this clock and hence the two endpoints of each interval $\Delta t_{\mathbf{O}}$ correspond to some object being brought into contact, through the action of $\hat{U}_{\mathbf{B_{O}}}$, with $\mathbf{O}$'s meter stick.  Hence $\hat{U}_{\mathbf{B_{O}}}$ can be regarded, without loss of generality, as a periodic function of $t_{\mathbf{O}}$ with period $\Delta t_{\mathbf{O}}$: with respect to some arbitrarily chosen initial value $t_{\mathbf{O}} = 0$, one or the other of $| 1 \rangle$ or $| 0 \rangle$ is observed at each subsequent time $t_{\mathbf{O}} = \Delta \mathnormal{t}_{\mathbf{O}}$, $t_{\mathbf{O}} = \mathrm{2} \Delta \mathnormal{t}_{\mathbf{O}}$, $t_{\mathbf{O}} = \mathrm{3} \Delta \mathnormal{t}_{\mathbf{O}}$, etc., so one or the other of $\hat{E}_{0}$ and $\hat{E}_{1}$ must act at each of these times.  Landauer's principle guarantees that the period $\Delta t_{\mathbf{O}}$ is finite, and hence that any such sequence requires finite observer-relative time.  Whether $\hat{E}_{0}$ or $\hat{E}_{1}$ acts so that $| 1 \rangle$ or $| 0 \rangle$ is observed after $N$ clock ticks, i.e. at $t_{\mathbf{O}} = \mathnormal{N \Delta t}_{\mathbf{O}}$ can, therefore, be viewed as determined by the values at $t_{\mathbf{O}} = \mathnormal{N \Delta t}_{\mathbf{O}}$ of two $t_{\mathbf{O}}$-dependent, anti-correlated phase angles $\phi_{1} (t_{\mathbf{O}})$ and $\phi_{2} (t_{\mathbf{O}})$.  As $\hat{U}_{\mathbf{B_{O}}}$ is unobservable, in principle, by $\mathbf{O}$, these phase angles must be unobservable, in principle, by $\mathbf{O}$.  They must, therefore, be represented as complex phases, i.e. there are real functions $\varphi_{1} (t_{\mathbf{O}})$ and $\varphi_{0} (t_{\mathbf{O}})$ such that $\phi_{1} (t_{\mathbf{O}}) = \imath \varphi_{1} (t_{\mathbf{O}})$ and $\phi_{1} (t_{\mathbf{O}}) = \imath \varphi_{1} (t_{\mathbf{O}})$.  Hence we can write:
\begin{equation}
\hat{U}_{\mathbf{B_{O}}} (\mathnormal{t_{\mathbf{O}}}) = \alpha_{\mathrm{0}} \mathnormal{e^{- \imath \varphi_{\mathrm{0}} (t_{\mathbf{O}})}} \hat{\mathnormal{E}}_{\mathrm{0}} + \alpha_{\mathrm{1}} \mathnormal{e^{- \imath \varphi_{\mathrm{1}} (t_{\mathbf{O}})}} \hat{\mathnormal{E}}_{\mathrm{1}}
\end{equation}
where $\alpha_{0}$ and $\alpha_{1}$ are real coefficients chosen so that $\alpha_{0}^{2} + \alpha_{1}^{2} = 1$.  As $\hat{E}_{0}$ and $\hat{E}_{1}$ resolve the identity on $\mathcal{H}_{\mathbf{B_{O}}}$, this $\hat{U}_{\mathbf{B_{O}}} (\mathnormal{t_{\mathbf{O}}})$ is unitary.  The physical dynamics represented by $\hat{U}_{\mathbf{B_{O}}}$ is, therefore, symmetric in $t_{\mathbf{O}}$.  From the ``god's eye'' perspective, this time symmetry corresponds to reversibility of the mechanism that brings objects into contact with the meter stick.  From $\mathbf{O}$'s perspective, it corresponds to $\mathbf{O}$'s inability to determine whether a bit string that can, in the absence of \textit{a priori} knowledge, only be regarded as random is being read forwards or backwards.  The unitarity of $\hat{U}_{\mathbf{B_{O}}}$ from $\mathbf{O}$'s perspective justifies the representation of the self-interaction $\hat{H}_{\mathbf{B_{O}}}$, again from $\mathbf{O}$'s perspective, as a Hamiltonian operator.

The unitarity of the state propagator $\hat{U}_{\mathbf{B_{O}}}$ for any closed, i.e. unobserved system $\mathbf{B_{O}}$ is a standard axiom of quantum theory (axiom ii of Zurek, 2003).  As before, if this axiom is false, decompositional equivalence must be violated.  A failure of unitarity is a failure of time invariance; it indicates that sequences of outcomes ordered in the $+t_{\mathbf{O}}$ direction can be distinguished, in practice, from sequences ordered in the $-t_{\mathbf{O}}$ direction.  As $t_{\mathbf{O}}$ is fully defined by the $\mathbf{O}$ - $\mathbf{B_{O}}$ partition, this asymmetry in $t_{\mathbf{O}}$ can only be a consequence of the definition of this partition.  Decompositional equivalence requires, however, that partitions and hence their definitions have no physical consequences; hence any such asymmetry violates decompositional equivalence.  Note that the existence of any observer-independent, ``objective'' time coordinate $t$ in $\mathbf{U}$ requires some boundary (e.g. an ``initial condition of the universe'') at which to be defined and similarly violates decompositional equivalence.  The (in principle unobservable) evolution of $\mathbf{U}$ is, therefore, time-symmetric for any arbitrarily-chosen, merely parametric coordinate $t$ and is hence unitary.  The unitarity of this inferred evolution $\hat{U}_{\mathbf{U}}$ justifies the assumption of an additively-decomposable Hamiltonian self-interaction $\hat{H}_{\mathbf{U}}$ as a representation of ``how the world works'' and hence justifies the representation of the $\mathbf{O}$ - $\mathbf{B_{O}}$ interaction as a Hamiltonian $\hat{H}_{\mathbf{OB_{O}}}$ for any arbitrarily chosen decomposition of $\mathbf{U}$ into $\mathbf{O}$ and $\mathbf{B_{O}}$.  

The objective asymmetry of time is a central feature of the classical worldview that is notoriously difficult to reproduce within quantum theory.  Objective, i.e. observer-independent decoherence is often suggested as a source of asymmetric time (e.g. Schlosshauer, 2007; Zurek, 2003), but as seen above, observer-independent decoherence requires the specification of an observer-independent boundary at which to act and thus violates decompositional equivalence as well as begging the question of classicality.  Rovelli (2014) has recently shown that an apparently asymmetric time, in any direction, can be obtained by suitably dividing a quantum universe into multiple ($N >> 2$) subsystems and then adopting a coarse-graining that defines certain variables as effectively classical.  This procedure is not only observer-relative but defines asymmetric time relative to a coarsely-observed subsystem; it therefore respects decompositional equivalence.

\subsection{The Born rule}

From the ``god's eye'' perspective, it is clear that the ratio $\alpha_{0}^{2} / \alpha_{1}^{2} = m / n$.  One would, moreover, expect from this perspective that over a sufficiently long elapsed time, the probabilities $P_{0}$ and $P_{1}$ of observing $| 0 \rangle$ and $| 1 \rangle$, respectively, would be such that $P_{0} / P_{1} = m / n$, i.e. one would expect the Born rule (axiom v in Zurek, 2003) to hold.  Because $\mathbf{B_{O}}$ is a black box, however, the numbers $m$ and $n$ cannot be determined by observation, so the equation $\alpha_{0}^{2} / \alpha_{1}^{2} = m/n$ cannot be derived by $\mathbf{O}$.  Similarly, the ergotic assumption that the observed frequencies of $| 0 \rangle$ and $| 1 \rangle$ between $t_{\mathbf{O}} = 0$ and $t_{\mathbf{O}} = \mathnormal{N \Delta t}_{\mathbf{O}}$ for some large $N$ correspond to probabilities of future observations cannot be proved valid for any black box, as its validity would contradict the Ashby-Moore theorem noted earlier.  The Born rule remains, therefore, a useful \textit{rule} for an observer, but it cannot be considered a theorem.  This pragmatic view of the Born rule has previously been advanced by Fuchs (2010) as the appropriate view within quantum Bayesianism.

\subsection{Hilbert space}

As the propagator $\hat{U}_{\mathbf{B_{O}}}$ is defined as acting on the state space $\mathcal{H}_{\mathbf{B_{O}}}$, $\mathbf{O}$ can attribute a state $| \mathbf{B_{O}} \rangle$ to $\mathbf{B_{O}}$ between observations by construction.  At the time of any observation, i.e. at $t_{\mathbf{O}} = \mathnormal{N \Delta t}_{\mathbf{O}}$ for any $N \geq 0$, this state is such that:

\begin{equation}
\hat{U}_{\mathbf{B_{O}}} (\mathnormal{t_{\mathbf{O}}}): | \mathbf{B_{O}} \mathnormal{(t_{\mathbf{O}} = N \Delta t_{\mathbf{O}})} \rangle \mapsto | \mathbf{B_{O}} \mathnormal{(t_{\mathbf{O}} = (N + \mathrm{1})} \mathnormal{\Delta t_{\mathbf{O}})} \rangle.
\end{equation}
Combining Eq. 3 and 4 to replace operators with the states they produce, we then have:

\begin{equation}
| \mathbf{B_{O}} \mathnormal{(t_{\mathbf{O}}) \rangle} = \alpha_{\mathrm{0}} \mathnormal{e^{- \imath \varphi_{\mathrm{0}} (t_{\mathbf{O}})} | \mathrm{0} \rangle} + \alpha_{\mathrm{1}} \mathnormal{e^{- \imath \varphi_{\mathrm{1}} (t_{\mathbf{O}})} | \mathrm{1} \rangle}
\end{equation}
at any $t_{\mathbf{O}}$.  

With the characterization of $| \mathbf{B_{O}} \mathnormal{(t_{\mathbf{O}})} \rangle$ given by Eq. 5, it is clear that the postulated $\mathcal{H}_{\mathbf{B_{O}}}$ is a finite dimensional and hence separable Hilbert space, that the observationally-distinguishable states $| 0 \rangle$ and $| 1 \rangle$ are basis vectors, and that $\hat{E}_{0}$ and $\hat{E}_{1}$ are von Neumann projections.  The generalization from two basis vectors to any finite number is, moreover, straightforward.  As any physically-implemented measurement device has finite resolution -- infinite resolution would violate Landauer's Principle -- infinite-dimensional Hilbert spaces are, as previously pointed out by Fuchs (2010), merely a convenience for performing calculations.

The requirement that $\mathcal{H}_{\mathbf{B_{O}}}$ is a separable Hilbert space is the final standard axiom of quantum theory (axiom i in Zurek, 2003).  Let us consider, once again, the physical consequences of a failure of this axiom.  First, were the state space of a physical system not separable, it would be impossible to enumerate the basis vectors and hence the physical degrees of freedom associated with any subsystem; the notion of ``decomposing'' the system into subsystems would then be ill-defined.  If the state space lacked an inner product, there would be no well-defined sense in which states could be regarded as similar or dissimilar.  A separable vector space is, therefore, required by the most basic intuitions regarding systems and their states.  On the other hand, strengthening the axiom by limiting the state space to real-valued coordinates is naturally interpreted as implying the in-principle observability of every state.  Unless additional axioms limiting the observability of some states or combinations of states are imposed, the represented system fails to be a black box, thus contradicting either decompositional equivalence or Landauer's principle.  As shown by Leifer (2014), ``hidden variable'' theories consistent with quantum theory must impose such additional axioms, either explicitly or implicitly through the definitions of the hidden variables; the observer's blindfold in Fig. 2 provides an example of such a constraint.  The space for which the \textit{observationally-distinguishable} states form a basis remains a Hilbert space in such theories, as shown here for the ``world'' of Fig. 2.

\subsection{Zurek's ``axiom o'' and tensor products}

Zurek (2003) includes among the axioms of quantum theory an ``axiom o'' stating that: 1) $\mathbf{U}$ consists of systems and 2) the Hilbert spaces of composite systems are tensor products of the Hilbert spaces of their constituents.  The second clause is sometimes (e.g. in Nielsen and Chaung, 2000) considered an independent axiom.  If a Hilbert-space representation of \textit{all} systems has already been assumed, this ``axiom'' serves to operationally define a ``constituent'' system; its effect, in this case, is to limit decompositions to finite numbers of constituents, thus avoiding non-separable infinite tensor products.

The first clause of ``axiom o'' can be read weakly as merely stating that decompositions of $\mathcal{H}_{\mathbf{U}}$ are possible, i.e. that the notion of a ``composite system'' makes sense.  It can also be read as the statement that all possible (finite) decompositions of $\mathcal{H}_{\mathbf{U}}$ objectively exist, i.e. exist independently of their observation by any observer.  Any reading of this axiom as postulating the objective existence of only some decompositions and not others, however, violates decompositional equivalence (Fields, 2014a).  Similarly, any reading of the axiom as postulating that all observers interact with exactly the same systems violates decompositional equivalence.  Either of these readings import into quantum theory a component of the classical worldview, that the universe comprises a collection of objectively well-defined, physically-bounded, time-persistent, fully observer-independent \textit{objects}, that the other axioms of quantum theory collectively contradict.

\subsection{Summary}

What has been shown here is that the primary axioms of quantum theory, and hence quantum theory itself, follow from the conjunction of decompositional equivalence and Landauer's principle.  If quantum theory is false, in particular, one or the other, or both, of these principles must be violated.  Both of these principles express deep physical intuitions: the intuition that there is an observer-independent way that the world works and the intuition that effort and therefore energy must be expended to acquire information.  We should, therefore, intuitively expect quantum theory to be true.  Indeed, we should be utterly astonished should we ever discover that quantum theory is not true. Should we ever discover, for example, that fundamental physical interactions change at the boundaries of macroscopic objects, our deepest intuitions about the world would have to be abandoned.

Landauer's principle and decompositional equivalence play complementary roles in quantum theory.  By imposing a finite energetic cost on information acquisition, Landauer's principle places a finite temporal interval between every pair of observations.  \textit{Every} observation is, therefore, coarse-grained in time as a matter of principle; finite resolution renders all measurements coarse-grained in the measured degree of freedom as well.  Coarse-graining in time gives the world an interval between observations in which to change its state.  Decompositional equivalence, on the other hand, restricts what can be observed; in particular, it denies observers access to internal boundaries and hence to bounded components of the world.  The only ``states'' that any observer $\mathbf{O}$ sees are states of that observer's world $\mathbf{B_{O}}$.  These states are distinguished and hence individuated only by observed outcome values, i.e. only by how the world affects the observer.  Between observations, they can only be represented as superpositions.  

By ruling out ``god's eye'' perspectives across the board, decompositional equivalence and Landauer's principle together generalize the restrictions on observational access noted for specific cases in Leifer (2014).  If the world is regarded as having, like the world depicted in Fig. 2, ``ontic'' states and behaviors that underlie the state changes observed by $\mathbf{O}$, these ontic states and behaviors of the world are unobservable in principle and hence unavailable as an empirical basis for theory construction.  The only states that are available as an empirical basis for theory construction are the observationally-distinguishable states, i.e. states labelled by distinct observed outcome values.  Such observationally-distinguishable states and the discrete transitions between them are fully described by quantum theory.

As decompositional equivalence and Landauer's principle together imply that any observer's ``world'' is a black box, the present results can be summarized as showing that the terms ``quantum system'' and ``black box'' are co-extensive.  Classical information theory can thus be regarded as a re-discovery of quantum theory from a different starting point and with a different notation.  Fuchs' use of the term ``interiority'' to characterize the essentially unpredictable nature of quantum systems (Fuchs, 2010) is therefore highly appropriate.

\section{The physical meaning of superposition}

\subsection{Systems and states}

The expressions in Eq. 3 and 5 are clearly superpositions; the operators $E_{0}$ and $E_{1}$ are superposed in Eq. 3 and the basis vectors $| 0 \rangle$ and $| 1 \rangle$ are superposed in Eq. 5.  The physicality of superpositions is widely regarded as one of the central mysteries of quantum theory.  It is, therefore, worth reflecting on what these expressions mean physically.  Doing so, as we will see, casts the standard no-go theorems of quantum theory in a new light, showing them to be the theory's way of forbidding the interpretation of state-space decompositions as defining observer-independent ``objects'' as they are conceived of within the classical worldview.

When the observationally-distinguishable states $|0\rangle$ and $|1\rangle$ are viewed as states of $\mathbf{B_{O}}$ as a single, whole system, the superposition $| \mathbf{B_{O}} \mathnormal{(t_{\mathbf{O}}) \rangle} = \alpha_{\mathrm{0}} \mathnormal{e^{- \imath \varphi_{\mathrm{0}} (t_{\mathbf{O}})} | \mathrm{0} \rangle} + \alpha_{\mathrm{1}} \mathnormal{e^{- \imath \varphi_{\mathrm{1}} (t_{\mathbf{O}})} | \mathrm{1} \rangle}$ (Eq. 5) has the straightforward meaning given in \S 5.7: the complex phases and hence the superposition itself reflect $\mathbf{O}$'s inability to observe the inner workings of $\mathbf{B_{O}}$ and hence inability to predict what $\mathbf{B_{O}}$ will do next.  This is the standard epistemic view of the quantum state developed by Bohr and Heisenberg and extended by many others into the ``$\psi$-epistemic'' interpretations current today (e.g. Spekkens, 2007; Bub and Pitowsky, 2010; Healey, 2012): it expresses $\mathbf{O}$'s objective ignorance of $\mathbf{B_{O}}$'s internal state, ignorance that cannot, even in principle, be resolved by observation.  

The system $\mathbf{B_{O}}$ does not, however, have to be regarded as a single, whole system.  Like any system, it can be arbitrarily decomposed.  One natural decomposition of $\mathbf{B_{O}}$ is into two systems $\mathbf{S}_{\mathrm{0}}$ and $\mathbf{S}_{\mathrm{1}}$, where $\mathbf{S}_{\mathrm{0}}$ is defined to be in state $|0 \rangle$ and $\mathbf{S}_{\mathrm{1}}$ is defined to be in state $|1 \rangle$.  This decomposition renders $\mathbf{B_{O}}$ a composite system.  It also renders the state of the composite system $\mathbf{B_{O}}$ unobservable: with this decomposition, $\mathbf{O}$ can observe either the state of $\mathbf{S}_{\mathrm{0}}$ or the state of $\mathbf{S}_{\mathrm{1}}$ but can never observe their joint state.  Letting $\mathbf{S} \mathnormal{(t_{\mathbf{O}})}$ be the notional single system occupying the states that $\mathbf{O}$ \textit{can} observe at $t_{\mathbf{O}}$ and replacing states with the systems that occupy them, Eq. 5 can be re-written:

\begin{equation}
\mathbf{S} \mathnormal{(t_{\mathbf{O}})} = \alpha_{\mathrm{0}} \mathnormal{e^{- \imath \varphi_{\mathrm{0}} (t_{\mathbf{O}})}} \mathbf{S}_{\mathrm{0}} + \alpha_{\mathrm{1}} \mathnormal{e^{- \imath \varphi_{\mathrm{1}} (t_{\mathbf{O}})}} \mathbf{S}_{\mathrm{1}}.
\end{equation}
Here the \textit{system} $\mathbf{S}$ is a superposition of the systems $\mathbf{S}_{\mathrm{0}}$ and $\mathbf{S}_{\mathrm{1}}$.  This superposition can be given an epistemic interpretation as above: it expresses $\mathbf{O}$'s objective ignorance regarding the system that will be observed next.

Suppose now that $\mathbf{O}$ has an additional observable; for example, suppose $\mathbf{O}$ can re-orient the meter stick vertically and receive a ``1'' or ``0'' outcome if an encountered object does or does not have a vertical dimension of 1 m.  In this case the ``system'' $\mathbf{S}_{\mathrm{1}}$ can be in one of two vertical ``states'': some 1 m wide objects are also 1 m high, while others are not (see Fig. 2).  The ``system'' $\mathbf{S}_{\mathrm{0}}$, however, can only be in one vertical state.  If, on the other hand, $\mathbf{O}$ is supposed to possess a shape observable that distinguishes objects by the numbers of corners they contain, both $\mathbf{S}_{\mathrm{0}}$ and $\mathbf{S}_{\mathrm{1}}$ can be in any of the three distinct ``states'' defined by having zero, three or four corners.  Nothing, moreover, prevents $\mathbf{O}$ from choosing to define ``systems'' by numbers of corners and to regard horizontal and vertical dimensions as ``states'' of those systems.  Decompositional equivalence allows any subset of the available observables to be regarded as defining systems, in which case the remaining subset of observables defines states of those systems.

If observers can obtain information only by observation -- i.e., if they occupy a universe satisfying decompositional equivalence -- the distinction between ``systems'' and ``states'' drawn above is the only such distinction available.  Decompositional equivalence explicitly forbids system boundaries, and hence the choices of observables that define them, from having any physical significance.  Human observers, for example, tend to use position, size and shape to define systems, e.g. to pick out their favorite chairs or voltmeters from among the other furniture of the laboratory.  If the universe satisfies decompositional equivalence, from the perspective of fundamental physics these choices of system-defining degrees of freedom are entirely arbitrary, and there is no principled reason to assume that other observers would make them.  

The lack of any physical distinction between observables that define systems and observables that characterize states has an important consequence: the claim that ``system $\mathbf{S}$ is in state $| \mathbf{S} \rangle$'' is simply the claim that some set of outcome values have been observed together.  These outcome values characterize the $\mathbf{O}$ - $\mathbf{B_{O}}$ interaction and hence the state $| \mathbf{B_{O}} \rangle$, entirely independently of any interpretative choices $\mathbf{O}$ may have made.  There is, therefore, no \textit{physical} sense in which $| \mathbf{B_{O}} \rangle$ is separable; $| \mathbf{B_{O}} \rangle$ remains a superposition as described by Eq. 5, or its generalization to some larger basis, at all times.  Entanglement is not, therefore, an occasional characteristic of some components of $\mathbf{B_{O}}$ but is rather, as Everett (1957) recognized, the permanent status of $\mathbf{B_{O}}$ itself.

\subsection{No-go theorems}

The non-objectivity of system boundaries required by decompositional equivalence corresponds, from an observer's perspective, to an unresolvable ambiguity in the source of any detected outcome.  In the world of Fig. 2, for example, $\mathbf{O}$ cannot determine whether a change in observed outcome from ``0'' to ``1'' reflects a change in the length of some single object that is in continuous contact with the meter stick or a change in which object has been placed in contact with the meter stick.  Similarly, $\mathbf{O}$ cannot determine whether a fixed outcome of ``0'' or ``1'' indicates continuous contact with a single object or an exchange of one object for another with the same length.  These ambiguities characterize the outcomes obtainable by any observer of any black box, regardless of the number of measurement devices that can be deployed or their resolution, as long as both are finite, i.e. as long as Landauer's Principle is respected.  

The well-known no-go theorems of quantum theory, including Bell's Theorem, the Kochen-Specker Theorem and the no-cloning theorem, follow from these ambiguities about the provenance of observational outcomes (for details, see Fields, 2013a).  Each of these theorems can be viewed as stating that a particular inference from observations is illegitimate; in doing so, each reveals assumptions that conflict with either Laudauer's principle, decompositional equivalence, or both.  The no-cloning theorem blocks the inference that the state of one system is a ``copy'' of the state of another system.  Two systems can be in the same state, however, only if the systems themselves are copies.  Decompositional equivalence prevents the access to system boundaries that would be required to establish this by finite observations.  The Kochen-Specker theorem blocks the inference that two distinct measurement procedures, both of which act on a single subset of degrees of freedom of a particular system, will produce the same sequence of outcomes.  Such an inference depends on the assumption that the outcomes of the measurements result \textit{only} from interactions with the specified degrees of freedom.  As noted earlier, decompositional equivalence prevents such exclusivity from being established by finite observations.  Bell's theorem blocks the inference that two measurements performed on distant parts of a system are independent; it shows, in particular, that local causation and counterfactually-definite states cannot both be assumed.  Decompositional equivalence and Landauer's principle together block the inference that \textit{any} two observations are causally independent.  They place the ``ontic'' causal structure of the world beyond observation, even in principle, and thus enforce a working assumption that this causal structure, whatever it is, is arbitrarily complex.  Even if the ontic causal structure of the world is classical as depicted in Fig. 2, the consequences of that structure for an observer can only be described, from the observer's own perspective, by quantum theory.

\subsection{Physics without systems?}

Physics is generally taken, by all but committed instrumentalists, to be about \textit{physical systems}.  Equations of motion, for example, are taken to describe how physical systems move.  Whenever a sentence begins with ``Let $\mathbf{S}$ be a system with Hilbert space $\mathcal{H}_{\mathbf{S}}$ ...'' or similar language, the existence of some physical system is being invoked.  Physical systems are, moreover, assumed in all but special cases both to persist through time and to be re-identifiable, at least in principle, through time.  Bohr's insistence on classical language for the description of laboratory tools and procedures is founded on a straightforwardly realist view that the laboratory itself and the apparatus, other equipment, and observers within it are all well-defined physical systems that unproblematically persist through time and are unproblematically re-identifiable following periods of non-observation.  The near-universal focus on \textit{states} as the subjects of superposition, entanglement, and other quantum effects in interpretations of quantum theory can be attributed to this deep faith that system identification is, or at least can be viewed as, unproblematic.

A rejection of the idea that ``physical systems'' can be unproblematically assumed can, however, be found already in Landsman's (2007) ``Stance 1'' about quantum theory, the claim that ``Quantum theory is fundamental and universally valid, and the classical world has only ``relative'' or ``perspectival'' existence'' (p. 422).  Nothing in this statement implies that effectively-classical \textit{states} have only relative existence - as they explicitly do in the formulations of Everett (1957) or Rovelli (1996), for example - while effectively-classical \textit{systems} such as experimental apparatus have a firmer, non-relative existence.  The discussions of superposition and the no-go theorems above only makes the natural equivalence of relative existence for states and relative existence for systems more explicit.  It is interesting, in this regard, that a view of systems as having only relative existence has recently emerged as a natural interpretative outcome of work in quantum cryptography, where the possibility that the device an observer thinks she is interacting with may not be the device that she is actually interacting with must be taken seriously.  As Grinbaum (2015) puts it, ``Device-independent methods convert the usually implicit trust of the observer (in, e.g. the identity of the system of interest or the experimental apparatus) into a theoretical problem.  By doing so, they erase one of the main dogmas of quantum theory: that it deals with systems'' (p. 2; parenthetical inserted).

When the ubiquitous superpositions in quantum theory are viewed as superpositions of systems, i.e. as indicating an unresolvable ambiguity in the sources of observational outcomes, they cease to be so mysterious.  While natural language expresses the conceit that ``systems'' are fixed and enduring things, we all know that this is not actually the case.  The microstructures of all systems are in constant flux.  Living systems obviously gain and lose material components, and hence become different \textit{physical systems}, on a daily if not hourly or even moment-to-moment basis; your own breathing provides an example.  It is, moreover, just an item of faith, sometimes explicitly propped up by appeals to Occam's razor, that systems do not undergo more radical alterations or even substitutions of one for another while we are not observing them.  The cognitive mechanisms underlying this item of faith are becoming better understood, largely by studying their points of failure (Fields, 2012; Fields, 2013b; Fields, 2014b).  Both cybernetics and quantum theory reminds us that this faith-driven inference of object identity through time is unsupportable, in principle, by finite observations.

\section{Quantum theory in practice: The double-slit experiment}

Quantum theory was originally developed not by considering thought experiments such as that presented here, but by searching for a mathematical formalism that could explain the results of otherwise-mystifying experiments.  Once it was understood that light behaved like a particle in the quantum domain, the already century-old double-slit experiment, which when performed by Thomas Young in 1803 had demonstrated that light behaved as a wave, became a mystery.  Indeed Feynman, in an often-quoted passage, describes it as ``a phenomenon which is impossible, \textit{absolutely} impossible, to explain in any classical way, and which has in it the heart of quantum mechanics. In reality, it contains the only mystery'' (Feynman, Leighton and Sands, 1965, Vol. III, Sect. 1.1, emphasis in original).  

The apparatus for the double-slit experiment is schematically illustrated, from the perspective of its designer, in Fig. 3a.  A source emits particles or waves of some kind: photons, electrons, C$_{60}$ ``Buckyballs'' (Arndt \textit{et al.}, 1999) or even heavier molecules with masses up to 10,000 amu (Eibenberger \textit{et al.}, 2013).  These pass though a pair of slits, the implementation of which depends on the type of particle/wave passing through them, and are detected at some distance beyond the slits.  These components of the apparatus are connected by a rigid structure that also isolates them, in darkness and under high vacuum, from the outside world.  Observers employing the double-slit apparatus to make measurements do not, therefore, have observational access to the internals of the instrument.  Observers interact, instead, with a ``user interface'' comprising controls that set the source intensity and the states, either open or closed, of the two slits, together with a display screen that shows the positions, on the $x-y$ plane perpendicular to the instrument's axis, of any detected particles.  The state of the instrument from the observer's perspective is defined by the states of these interface elements, as shown in Fig. 3b.

\begin{figure}
\centering
\includegraphics[width=160mm]{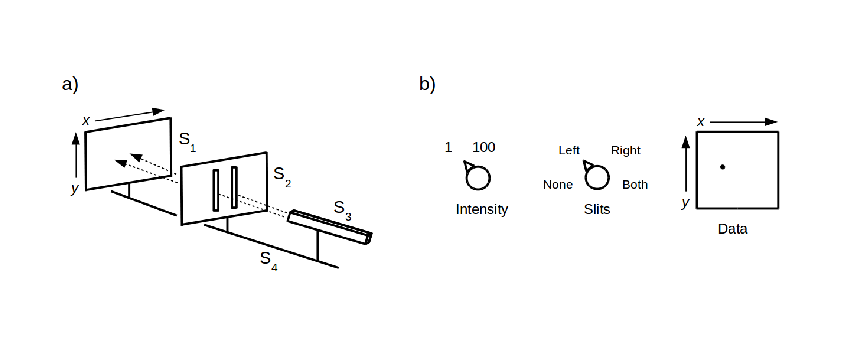}
\caption{a) Schematic representation of the double-slit apparatus from the perspective of its designer, comprising a detection screen $\mathbf{S_{1}}$, slits $\mathbf{S_{2}}$, photon or ion source $\mathbf{S_{3}}$ and stabilizing structure $\mathbf{S_{4}}$.  Real instruments may be substantially more complex, particularly in the implementation of the slits.  The dashed arrows show classical trajectories used to collimate the apparatus.  b) The double-slit apparatus from the perspective of an observer, comprising a knob for regulating the source intensity, a knob for determining which slit(s) to open, and a display screen on which detection events are recorded.  The $x$ and $y$ coordinates of the display correlate with the $x$ and $y$ coordinates of the detector as shown.}
\end{figure}

The phenomenology of the double-slit experiment is straightforward.  If neither slit is open, nothing is detected on the screen.  If only one slit is open, particles are detected in an approximately Gaussian distribution, elongated on the $y$ axis and centered on the $x-y$ position of the open slit.  If both slits are open, an interference pattern appears along the $x$ axis of the display, with the interference peak and valley intensities indicated on the $y$ axis.  These patterns are maintained even if the source intensity is set to just one particle per unit time.  If, however, the detection screen is moved to be immediately behind the slits (for photons) or some means of detecting which slit a particle passed through is installed (for massive particles), the interference pattern disappears.  The standard interpretation of this phenomenology is that individual particles, even individual large molecules such as the porphyrins employed by Eibenberger \textit{et al.} (2013), act as waves, passing through both slits if both are open and subsequently interfering with themselves.  Both this ambiguous wave-particle behavior and the fact that the interference pattern disappears if the particle's trajectory is ``observed'' by monitoring the slit it passed through are canonical ``quantum'' phenomena.  

Let us now consider this standard interpretation in somewhat more detail.  If the source intensity is adjusted so that many particles are emitted per unit time, the resulting ``beam'' of particles illuminates the plate $\mathbf{S_{2}}$ in which the slits are cut.  Some fraction of the ``beam'' passes through the slits and subsequently, the ``beam'' components that passed through the left slit interfere with the components that passed through the right slit.  This situation is analogous to what happens when water waves encounter a breakwater barrier with two openings, as often demonstrated in laboratory exercises for beginning physics students.  Hence the ``beam'' behaves like a classical wave.  This classical analogy breaks down, however, when the source intensity is turned down to one particle per unit time.  As the interference pattern persists, built up one detection event at a time, it must be assumed either that each particle went through both slits and subsequently interfered with itself, or else that the particles can interfere with each other across time, with the first assumption by far the most commonly chosen.  In this case the trajectory of each particle can be thought of as comprising two components.  First, the particle travels from the source to both slits; representing the $x$ coordinates of the slits as $l$ and $r$ and ignoring the other coordinates, this first part of the trajectory can be represented as $| l + (r - l)/2 \rangle$ (the $x$ coordinate location of the source) $\rightarrow (1/ \sqrt{2})(|l \rangle + |r \rangle)$.  This first part of the trajectory has, in other words, a fixed starting point but a superposition of end points.  The second part of the trajectory can then be represented as $(1/ \sqrt{2})(|l \rangle + |r \rangle) \rightarrow |x_{d} \rangle$, where $x_{d}$ is the $x$ coordinate of the detection event.  Each particle is, therefore, considered to be in a superposition of $x$ coordinate states for its entire trajectory, excepting only its initial location at the source and its final location on the detector.

The $x$ coordinate of each particle is, however, \textit{observed} only at the detector.  No aspect of its trajectory is observed other than its detection at $S_{1}$.  Indeed, that the particle even has a trajectory is an interpretative assumption based on the designer's, not the observer's, perspective on the apparatus.  The discussion in the preceding sections tells us that, from the observer's perspective, the apparatus is a black box, and it warns us that assumptions made from other perspectives can rapidly lead to contradictions.  Let us, therefore, set all such assumptions aside and focus exclusively on the perspective of the observer.  What the observer can determine by observation is that the number of detection events correlates with the source intensity provided that at least one slit is open, and that the number of open slits determines the pattern - either Gaussian or interference - that is detected.  By appeal to the classical theory of waves, the observer can infer that the slits are far enough from the detection screen for an interference pattern to develop whenever both slits are open.  Where, however, is the source?  It is consistent with the observed detection patterns that each slit is also a source; if a slit is open, it emits particles at the chosen intensity, but if it is closed it does not.   The question: ``What is the source of a particle detected at time $t$?'' has, therefore, a determinate answer if only one slit is open, but if both slits are open, the answer is ambiguous.  This ambiguity can be expressed by representing the source as being in a superposition $(1/\sqrt{2})(|l \rangle + |r \rangle)$ of positions along the $x$ axis, or following the reasoning in \S 6, by representing the source as a superposition of systems $(1/\sqrt{2})(\mathbf{S_{3\mathnormal{l}}} + \mathbf{S_{3\mathnormal{r}}})$, where the notations ``$\mathbf{S_{3\mathnormal{l}}}$'' and ``$\mathbf{S_{3\mathnormal{r}}}$'' refer to left and right sources of particles.  These alternative expressions equally and completely capture what an observer can know about the source of an observed particle.  No more detail can be given.  The question, ``Where was the particle before it interacted with the slits'' cannot be properly posed.

It is important to note that taking the apparatus apart to allow inspection of the internal components does not resolve the ambiguity.  Once the instrument is disassembled, the interference pattern disappears.  Indeed any attempt to determine whether the particle has a trajectory prior to encountering the slits disrupts the pattern, just as attempting to determine which slit has been traversed does.  The source superpositions $(1/\sqrt{2})(|l \rangle + |r \rangle)$ or $(1/\sqrt{2})(\mathbf{S_{3\mathnormal{l}}} + \mathbf{S_{3\mathnormal{r}}})$, and the interference pattern they produce, can therefore be viewed as a consequence - perhaps even a side-effect - of the isolation and consequent unobservability of the internals of the apparatus.  They are, in other words, consequences of the apparatus being a black box.  

Viewing the double-slit experiment in this way enables a clear prediction: interference patterns should be observable in \textit{any} situation in which the epistemic position of the observer is as represented in Fig. 3b.  From this perspective, the steadily-increasing reports of interference phenomena in studies of human categorization that access information from multiple locations in the category network (reviewed by Pothos and Busemeyer, 2013; Aerts, Gabora and Sozzo, 2013) are not surprising.

\section{Conclusion}

A scientific theory is a collection of intuitions and assumptions around which a mathematical formalism that enables precise and testable predictions has been constructed.  Whether the formalism is fully consistent with the intuitions and assumptions on which it is based, and with the support of which it is used to generate predictions, may not be obvious.  Quantum theory provides a dramatic case in point: the formal axioms of quantum theory are notoriously non-intuitive, and the extent to which they are consistent with common physical intuitions and assumptions is not obvious.  The unexpected experimental observations that motivated quantum theory, the no-go theorems, and the continuing demonstrations of entanglement and other distinctly ``quantum'' phenomena all indicate that some, at least, of the intuitions and assumptions that constitute the ``classical worldview'' are wrong.  It has often been argued that it is the implicit addition of such wrong assumptions that render quantum theory paradoxical (e.g. Peres and Fuchs, 2000).  The challenge is to determine precisely which assumptions are wrong.

What has been shown here is that two deep intuitions, that there is an observer-independent way the world works and that acquiring information requires spending energy, are consistent with quantum theory, and when formulated more precisely as the principle of decompositional equivalence and Landauer's principle, together materially imply quantum theory.  Both decompositional equivalence and Landauer's principle place limits on the information that can be obtained by observation; quantum theory provides a formalism for describing observations of the physical world that enforces these limits.  Empirical evidence supporting quantum theory is empirical evidence that these limits are real.  It is also evidence that all intuitions or assumptions that conflict with decompositional equivalence or Landauer's principle are wrong.  Combining intuitions or assumptions that conflict with either of these principles with quantum theory is guaranteed to lead to paradoxes and inconsistencies.  When such intuitions or assumptions are rigorously removed from quantum theory, the paradoxes can be expected to disappear.

Several intuitions or assumptions that play central roles in the classical worldview clearly do conflict with decompositional equivalence, Landauer's principle, or both.  These include:

\begin{itemize}
\item That observers can interact exclusively with particular, identified subsystems of the world.
\item That observational outcomes can be exclusively attributed to particular, identified subsystems of the world.
\item That observers can ``prepare'' two or more particular, identified subsystems of the world independently of each other.
\item That multiple observers can reliably identify and exclusively interact with the same subsystem of the world.
\item That observers can reliably identify and exclusively interact with the same subsystem of the world at multiple times.
\end{itemize}
These assumptions all reflect underlying intuitions that \textit{apparent subsystem boundaries are objectively real} and that \textit{upper limits can be placed on the complexity of the world}.  These deep intuitions directly contradict decompositional equivalence and Landauer's principle, respectively.  They therefore contradict quantum theory.  

It is important, moreover, to emphasize that by contradicting decompositional equivalence and Landauer's principle, these assumptions also contradict \textit{classical} physics, in particular, the associativity of classical phase space decomposition, the additivity of the classical Hamiltonian, and the 2nd Law of Thermodynamics.  While classical physics is often taken, informally, to both support and embody the classical worldview, nothing in classical physics requires or even suggests that any of the above five statement are true.  In his famous remark about the moon, Einstein clearly assumes that his hypothetical observer would be able to re-identify the moon as being the very same physical system seen before looking away; however, nothing in classical physics requires or even suggests that observers have such capabilities.  Indeed classical cybernetics, a discipline firmly based on classical physics, in particular on classical thermodynamics, demonstrates as discussed above that none of these assumptions can be true of any observers that are limited to finite means.

Because the intuitions or assumptions comprising the classical worldview are taken for granted in ordinary life, they appear regularly in foundational discussions of quantum theory.  The near 90-year project of interpreting quantum theory can, indeed, be regarded as the project of showing that not just the \textit{laws of classical physics} but also the \textit{classical worldview itself} emerge as approximations from or are otherwise somehow justified by quantum theory.  This is evident even in Landsman's (2007) formulation of his three categorical ``stances'' regarding the interpretation of quantum theory, all of which associate positions on the correctness or completeness of quantum \textit{theory} with the metaphysical status - ``absolute existence'' or not - of the classical \textit{world}.  Landsman's gloss of ``the classical world'' as ``what observation shows us to behave - with appropriate accuracy -
according to the laws of classical physics'' (p. 423) implicitly assumes, as Landsman later points out, a commonsense-realist metaphysics, one that nothing in classical physics requires.  

This project to produce the classical worldview from quantum theory is, moreover, required by most of its adherents to be either independent of observers altogether or independent of any particular characterization of observers.  Zurek (2003), for example, states explicitly that ``the observer's mind (that verifies, finds out, etc.) constitutes
a primitive notion which is prior to that of scientific reality'' (p. 363-364).  Hartle (2011) insists that the information obtainable by observers is ``a feature of the universe independent of human cognition or decision'' (p. 983) and hence, apparently, independent of the observational methods or tools available to the observers.  While some (e.g. von Neumann, 1932; Wigner, 1962) require observers to be conscious, most adopt the position of Rovelli (1996) or Schlosshauer (2007) that the observer is simply a physical system.  Fuchs (2010) lampoons the idea that quantum theory should include an explicit model of the observer: ``Would one ever imagine that the notion of an agent, the user of the theory, could be derived out of its conceptual apparatus?'' (p. 8).  The assumptions of the classical worldview are, therefore, generally treated by the interpretative project not just as facts, but as observer-independent facts.  As noted earlier, Zurek's ``axiom o'' can easily be construed as a statement that apparent subsystem boundaries are objectively real, and in this form it underlies the assumption of an objective, observer-independent system - environment boundary in decoherence calculations.  The assumption that observers can reliably identify and exclusively interact with the same subsystem of the world at multiple times underlies the frequentist conception of probability.  The recent Pusey-Barrett-Rudolph theorem (for an extensive review, see Leifer, 2014) relies on the assumption that spatially-separated systems can be prepared independently by either the same or different observers.  In discussing this assumption, Leifer remarks that ``we generally think that experiments on separated systems are independent of one another ... If we allow that genuinely global properties may be relevant even to an isolated system then we open up a Pandora's box'' (Leifer, 2014, p. 104).  If the global state of the world is non-separable, there are no isolated systems.  Pandora's box is open and cannot be closed.  As Galileo put it so long ago, Nature is indifferent to the difficulties that this might cause us.

Both decompositional equivalence and Landauer's principle are straightforwardly realist physical principles: they are claims about how a real, physical universe works.  As shown here, however, their consequences for observers are deeply if not radically instrumentalist.  It could be argued that this level of instrumentalism makes science impossible.  Perhaps it does make a certain na\"{i}ve-realist ideal of science impossible, but a pragmatic, everyday science is obviously not just possible but extraordinarily successful.  \textit{What makes our pragmatic science possible} becomes, in the present framework, a deep empirical question.  How, for example, are we able to identify systems of interest through time, isolate them ``well enough'' to conduct experiments, and agree among ourselves that we have replicated both the system of interest and the experimental manipulations?  How are we able to treat the world as a substantially white box instead of a completely black one?  These are not, it is worth emphasizing, philosophical questions to be resolved by either metaphysical speculation or interpretative analysis.  They are straightforwardly empirical questions that, \textit{contra} Fuchs (2010), demand experimental investigation and empirical theory-building.  The fact that we can get away, in most ordinary circumstances, with pragmatic assumptions that violate decompositional equivalence and/or Landauer's principle can only be an outcome of the way the world works, including the way that we as observers happen to be embedded in it.  It is an empirical challenge for physics to understand this happy outcome.

\section*{Acknowledgements}

Thanks to Don Hoffman for encouraging me to think about 1-bit information transfers, to The Federico and Elvia Faggin Foundation for financial support during the final stages of this work, and to an anonymous referee for suggestions and an additional reference.

\section*{References}

\hangindent=1cm 
Aerts, D., Gabora, L. and Sozzo, S. (2013).  Concepts and their dynamics: A quantum-theoretic modeling of human thought.  {\em Topics in Cognitive Science} 5: 737-772.

\hangindent=1cm 
Anders, J., Shabbir, S., Hilt, S. and Lutz, E. (2006).  Landauer's principle in the quantum domain.  {\em Electronic Proceedings in Theoretical Computer Science} 26: 13-18.

\hangindent=1cm
Arndt, M., Nairz, O., Vos-Andreae, J., Keller, C., van der Zouw, G. and Zeilinger, A. (1999).  Wave-particle duality of C$_{60}$ molecules.  {\em Nature} 401: 680-682.

\hangindent=1cm 
Ashby, W. R. (1956).  {\em An Introduction to Cybernetics}.  London: Chapman and Hall.

\hangindent=1cm 
Bacciagaluppi, G. and Valentini, A. (2009).  {\em Quantum Theory at the Crossroads: Reconsidering the 1927 Solvay Conference.}  Cambridge: Cambridge University Press.

\hangindent=1cm 
Bastin, T. (Ed.) (1971). {\em Quantum Theory and Beyond.} Cambridge: Cambridge University Press.  re-published by Cambridge University Press, 2009.

\hangindent=1cm
Bennett, C. H. (2003).  Notes on Landauer's principle, reversible computation, and Maxwell's Demon.  {\em Studies in the History and Philosophy of Modern Physics} 34: 501-510.

\hangindent=1cm 
Bub, J. and Pitowsky, I. (2010).  Two dogmas about quantum mechanics.  In: Saunders, S., Barrett, J., Kent, A. and Wallace, D. (Eds)  {\em Many Worlds? Everett, Quantum Theory and Reality}.  Oxford: Oxford University Press (pp. 433-459).

\hangindent=1cm 
Bunge, M. (1956).  Survey of the interpretations of quantum mechanics.  {\em American Journal of Physics} 24: 272-286.

\hangindent=1cm 
Butterfield, J. (2011).  Emergence, reduction and supervenience: A varied landscape.  {\em Foundations of Physics} 41: 920-959.

\hangindent=1cm 
Cabello, A. (2015).  Interpretations of quantum theory: A map of madness.  Preprint arXiv:1509.04711v1 [quant-ph].

\hangindent=1cm 
DeWitt, B. S. (1970).  Quantum mechanics and reality: Could the solution to the dilemma of indeterminism be a universe in which all possible outcomes of an experiment actually occur?  {\em Physics Today} 23(9): 30-40.

\hangindent=1cm
Eibenberger, S., Gerlich, S., Arndt, M., Mayor, M. and T\"uxen J. (2013).  Matter-wave interference of particles selected from a molecular library with masses exceeding 10,000 amu.  {\em Physical Chemistry and Chemical Physics} 15: 14696-14700.  

\hangindent=1cm 
Everett, H. III (1957).  ``Relative state'' formulation of quantum mechanics.  {\em Reviews of Modern Physics} 29: 454-462.

\hangindent=1cm 
Feynman, R. P., Leighton, R. B. and Sands, M. (1965).  {\em Feynman Lectures on Physics}.  New York: Addison-Wesley.

\hangindent=1cm 
Fields, C. (2011).  Classical system boundaries cannot be determined within quantum Darwinism.  {\em Physics Essays} 24: 518-522.

\hangindent=1cm 
Fields, C. (2012).  The very same thing: Extending the object-token concept to incorporate causal constraints on individual identity.  {\em Advances in Cognitive Psychology} 8: 234-247.

\hangindent=1cm 
Fields, C. (2013a).  Bell's Theorem from Moore's Theorem.  {\em International Journal of General Systems} 42: 376-385.

\hangindent=1cm 
Fields, C. (2013b).  The principle of persistence, Leibniz's Law and the computational task of object identification.  {\em Human Development} 56: 147-166.

\hangindent=1cm 
Fields, C. (2014a).  On the Ollivier-Poulin-Zurek definition of objectivity.  {\em Axiomathes} 24: 137-156.

\hangindent=1cm 
Fields, C. (2014b).  Motion, identity and the bias toward agency.  {\em Frontiers in Human Neuroscience} 8: Article \# 597.  

\hangindent=1cm
Fuchs, C. A. (2002).  Quantum mechanics as quantum information (and only a little more).  Preprint arXiv:quant-ph/0205039v1.

\hangindent=1cm
Fuchs, C. A. (2010).  QBism: The perimeter of quantum Bayesianism.  Preprint arXiv:1003.5209v1 [quant-ph].

\hangindent=1cm
Grinbaum, A. (2015).  How device-independent approaches change the meaning of Physics.  Preprint arXiv:1512.01035v1.

\hangindent=1cm
Ghirardi, G. C., Rimini, A. and Weber, T. (1986).  Unified dynamics for microscopic and macroscopic systems.  {\em Physical Review D} 34: 470-491.

\hangindent=1cm
Hartle, J. B. (2011).  The quasiclassical realms of this quantum universe.  {\em Foundations of Physics} 41: 982-1006.

\hangindent=1cm
Healey, R. (2012).  Quantum theory: A pragmatist approach.  {\em British Journal for The Philosophy of Science} 63: 729-771.

\hangindent=1cm
Jauch, J. M. (1968).  {\em Foundations of Quantum Mechanics.}  Reading, MA: Addison-Wesley.

\hangindent=1cm
Jennings, D. and Leifer, M. (2016).  No return to classical reality.  {\em Contemporary Physics} 57(1): 60-82.

\hangindent=1cm 
Kastner, R. (2014).  `Einselection' of pointer observables: The new H-theorem? {\em Studies in the History and Philosophy of Modern Physics} 48: 56-58.

\hangindent=1cm
Koenderink, J. (2014).  The all-seeing eye?  {\em Perception} 43(1): 1-6.

\hangindent=1cm 
Landauer, R. (1961).  Irreversibility and heat generation in the computing process.  {\em IBM J. Research Development}  5(3): 183-195.

\hangindent=1cm 
Landauer, R. (1999).  Information is a physical entity.  {\em Physica A} 263: 63-67.

\hangindent=1cm 
Landsman, N. P. (2007).  Between classical and quantum.  In: Butterfield, J. and Earman, J. (Eds)  {\em Handbook of the Philosophy of Science: Philosophy of Physics.}  Amsterdam: Elsevier (pp. 417-553).

\hangindent=1cm 
Leifer, M. S. (2014).  Is the quantum state real?  An extended review of $\psi$-ontology theorems.  {\em Quanta} 3: 67-155.

\hangindent=1cm 
Moore, E. F. (1956).  Gedankenexperiments on sequential machines.  In: Shannon, C. W. and McCarthy, J. (Eds) {\em Automata Studies}. Princeton, NJ: Princeton University Press, pp. 129-155.

\hangindent=1cm 
Nagel, E. (1961).  {\em The Structure of Science: Problems in the Logic of Scientific Explanation}.  New York: Harcourt, Brace \& World.

\hangindent=1cm
Nielsen, M. A. and Chaung, I. L. (2000).  {\em Quantum Information and Quantum Computation.}  Cambridge, UK: Cambridge University Press.

\hangindent=1cm
Norsen, T. and Nelson, S. (2013).  Yet another snapshot of foundational attitudes toward quantum mechanics.  Preprint arXiv:1306.4646v2 [quant-ph]. 
  
\hangindent=1cm
Ollivier, H., Poulin, D. and Zurek, W. H. (2004). Objective properties from subjective quantum states: Environment as a witness.  {\em Physical Review Letters} 93: 220401.

\hangindent=1cm
Ollivier, H., Poulin, D. and Zurek, W. H. (2005). Environment as a witness: Selective proliferation of information and
emergence of objectivity in a quantum universe.  {\em Physical Review A} 72: 042113.

\hangindent=1cm
Penrose, R. (1996).  On gravity's role in quantum state reduction.  {\em General Relativity and Gravitation} 28: 581-600.

\hangindent=1cm
Peres, A. and Fuchs, C. (2000).  Quantum theory needs no ``interpretation.''  {\em Physics Today} 53(3): 70-73.

\hangindent=1cm
Peres, A. and Terno, D. R. (2004).  Quantum information and relativity theory.  {\em Reviews of Modern Physics} 76: 93-123.

\hangindent=1cm
Pothos, E. M. and Busemeyer, J. M. (2013).  Can quantum probability provide a new direction for cognitive modeling?
{\em Behavioral and Brain Sciences} 36: 255-327.

\hangindent=1cm
Rovelli, C. (1996).  Relational quantum mechanics.  {\em International Journal of Theoretical Physics} 35: 1637-1678.

\hangindent=1cm
Rovelli, C. (2014).  Why do we remember the past and not the future?  The `time oriented coarse graining' hypothesis.  Preprint arXiv 1407.3384v2.

\hangindent=1cm
Shannon, C. E. (1948).  A mathematical theory of communication.  {\em Bell System Technical Journal} 27: 379-423.

\hangindent=1cm
Schlosshauer, M. (2007).  {\em Decoherence and the Quantum to Classical Transition.} Berlin: Springer.

\hangindent=1cm
Schlosshauer, M., Kofler, J. and Zeilinger, A. (2013).  A snapshot of foundational attitudes toward quantum mechanics.  {\em Studies in the History and Philosophy of Modern Physics} 44: 222-230.

\hangindent=1cm
Spekkens, R. W. (2007).  Evidence for the epistemic view of quantum states: A toy theory.  {\em Physical Review A} 75: 032110.

\hangindent=1cm
Sommer, C. (2013). Another survey of foundational attitudes towards quantum mechanics.  Preprint arXiv:1303.2719v1 [quant-ph]

\hangindent=1cm
Tegmark, M. (2000).  Importance of quantum decoherence in brain processes.  {\em Physical Review E} 61: 4194-4206.

\hangindent=1cm
Tegmark, M. (2012).  How unitary cosmology generalizes thermodynamics and solves the inflationary entropy problem.
{\em Physical Review D} 85: 123517.

\hangindent=1cm
von Neumann, J. (1932).  {\em Mathematical Foundations of Quantum Theory}  Berlin: Springer.

\hangindent=1cm 
Wallace, D. (2008).  Philosophy of quantum mechanics. In: Rickles, D. (Ed.)  {\em The Ashgate Companion to
Contemporary Philosophy of Physics.}  Aldershot, U.K.: Ashgate (pp. 16-98).

\hangindent=1cm
Wang, Q., Schoenlein, R. W., Peteanu, L. A., Mathies, R. A. and Shank, C. V. (1994).  Vibrationally coherent photochemistry in the femtosecond primary event of vision.  {\em Science} 266: 422-424.

\hangindent=1cm
Weinberg, S. (2012).  Collapse of the state vector.  {\em Physical Review A} 85: 062116.

\hangindent=1cm
Wigner, E. P. (1962).  Remarks on the mind-body question.  In: Good, I. J. (Ed.)  {\em The Scientist Speculates}.  New York: Basic Books (pp. 284-301).

\hangindent=1cm 
Zeh, D. (1970). On the interpretation of measurement in quantum theory.  {\em Foundations of Physics} 1: 69-76.

\hangindent=1cm
Zurek, W. H. (2003).  Decoherence, einselection, and the quantum origins of the classical.  {\em Reviews of Modern Physics} 75: 715-775.

\end{document}